%% file: main.tex
\pdfoutput=1

\documentclass[12pt,a4paper]{article}

\usepackage{multirow}

\usepackage{ifthen} 
\newboolean{pdflatex}
\setboolean{pdflatex}{true} 

\newboolean{articletitles}
\setboolean{articletitles}{true} 

\newboolean{uprightparticles}
\setboolean{uprightparticles}{false} 

\newboolean{inbibliography}
\setboolean{inbibliography}{false} 

\input{preamble}
\usepackage{longtable} 

\begin{document}

\renewcommand{\thefootnote}{\fnsymbol{footnote}}
\setcounter{footnote}{1}

\input{title-LHCb-PAPER}


\renewcommand{\thefootnote}{\arabic{footnote}}
\setcounter{footnote}{0}


\pagestyle{plain} 
\setcounter{page}{1}
\pagenumbering{arabic}


\input{introduction}

\input{detector}

\input{selection}

\input{physicsBg}

\input{fit}

\input{efficiency}
\input{systematics}

\input{summary}

\input{acknowledgements}


\addcontentsline{toc}{section}{References}
\bibliographystyle{LHCb}
\bibliography{main,LHCb-PAPER,LHCb-CONF,LHCb-DP}

\end{document}

%% file: preamble.tex
\usepackage{microtype}
\usepackage{lineno}  
\usepackage{xspace} 

\usepackage{graphicx}  
\usepackage{color}
\usepackage{colortbl}
\graphicspath{{./figs/}} 

\usepackage{amsmath} 
\usepackage{amssymb}
\usepackage{amsfonts}
\usepackage{upgreek} 

\newcommand*\patchAmsMathEnvironmentForLineno[1]{%
\expandafter\let\csname old#1\expandafter\endcsname\csname #1\endcsname
\expandafter\let\csname oldend#1\expandafter\endcsname\csname
end#1\endcsname
 \renewenvironment{#1}%
   {\linenomath\csname old#1\endcsname}%
   {\csname oldend#1\endcsname\endlinenomath}%
}
\newcommand*\patchBothAmsMathEnvironmentsForLineno[1]{%
  \patchAmsMathEnvironmentForLineno{#1}%
  \patchAmsMathEnvironmentForLineno{#1*}%
}
\AtBeginDocument{%
\patchBothAmsMathEnvironmentsForLineno{equation}%
\patchBothAmsMathEnvironmentsForLineno{align}%
\patchBothAmsMathEnvironmentsForLineno{flalign}%
\patchBothAmsMathEnvironmentsForLineno{alignat}%
\patchBothAmsMathEnvironmentsForLineno{gather}%
\patchBothAmsMathEnvironmentsForLineno{multline}%
}

\input{lhcb-symbols-def} 

\usepackage{cite} 
\usepackage{mciteplus}

%% file: lhcb-symbols-def.tex
\def\lhcb {\mbox{LHCb}\xspace}

\ifthenelse{\boolean{uprightparticles}}%
{

 \def\Pmu         {\ensuremath{\upmu}\xspace}

 \def\Ppi         {\ensuremath{\uppi}\xspace}

 \def\Ppsi        {\ensuremath{\uppsi}\xspace}

 \def\PDelta      {\ensuremath{\Delta}\xspace}                 
 \def\PXi      {\ensuremath{\Xi}\xspace}                 
 \def\PLambda      {\ensuremath{\Lambda}\xspace}                 
 \def\PSigma      {\ensuremath{\Sigma}\xspace}                 
 \def\POmega      {\ensuremath{\Omega}\xspace}                 
 \def\PUpsilon      {\ensuremath{\Upsilon}\xspace}                 
 
 \def\PB      {\ensuremath{\mathrm{B}}\xspace}                 
                  
 \def\PD      {\ensuremath{\mathrm{D}}\xspace}

 \def\PJ      {\ensuremath{\mathrm{J}}\xspace}                 
 \def\PK      {\ensuremath{\mathrm{K}}\xspace}

 \def\PW      {\ensuremath{\mathrm{W}}\xspace}

 \def\Pb      {\ensuremath{\mathrm{b}}\xspace}                 
 \def\Pc      {\ensuremath{\mathrm{c}}\xspace}

 \def\Pi      {\ensuremath{\mathrm{i}}\xspace}

 \def\Pp      {\ensuremath{\mathrm{p}}\xspace}

 \def\Ps      {\ensuremath{\mathrm{s}}\xspace}

}
{

 \def\Pmu         {\ensuremath{\mu}\xspace}

 \def\Ppi         {\ensuremath{\pi}\xspace}

 \def\Ppsi        {\ensuremath{\psi}\xspace}                 
                  
 \mathchardef\PDelta="7101
 \mathchardef\PXi="7104
 \mathchardef\PLambda="7103
 \mathchardef\PSigma="7106
 \mathchardef\POmega="710A
 \mathchardef\PUpsilon="7107
                  
 \def\PB      {\ensuremath{B}\xspace}                 
                  
 \def\PD      {\ensuremath{D}\xspace}

 \def\PJ      {\ensuremath{J}\xspace}                 
 \def\PK      {\ensuremath{K}\xspace}

 \def\PW      {\ensuremath{W}\xspace}

 \def\Pb      {\ensuremath{b}\xspace}                 
 \def\Pc      {\ensuremath{c}\xspace}

 \def\Pi      {\ensuremath{i}\xspace}

 \def\Pp      {\ensuremath{p}\xspace}

 \def\Ps      {\ensuremath{s}\xspace}

}

\def\mumu       {\ensuremath{\Pmu^+\Pmu^-}\xspace}

\def\Wpm    {\ensuremath{\PW^\pm}\xspace}

\def\squark    {\ensuremath{\Ps}\xspace}

\def\cquark    {\ensuremath{\Pc}\xspace}

\def\bquark    {\ensuremath{\Pb}\xspace}

\def\pion  {\ensuremath{\Ppi}\xspace}

\def\pip   {\ensuremath{\pion^+}\xspace}
\def\pim   {\ensuremath{\pion^-}\xspace}

\def\kaon  {\ensuremath{\PK}\xspace}
  \def\Kbar  {\kern 0.2em\overline{\kern -0.2em \PK}{}\xspace}

\def\KS    {\ensuremath{\kaon^0_{\rm\scriptscriptstyle S}}\xspace}

  \def\Dbar    {\kern 0.2em\overline{\kern -0.2em \PD}{}\xspace}

\def\B       {\ensuremath{\PB}\xspace}
\def\Bbar    {\ensuremath{\kern 0.18em\overline{\kern -0.18em \PB}{}}\xspace}

\def\Bz      {\ensuremath{\B^0}\xspace}

\def\jpsi     {\ensuremath{{\PJ\mskip -3mu/\mskip -2mu\Ppsi\mskip 2mu}}\xspace}
\def\psitwos  {\ensuremath{\Ppsi{(2S)}}\xspace}

  \def\Y#1S{\ensuremath{\PUpsilon{(#1S)}}\xspace}

\def\proton      {\ensuremath{\Pp}\xspace}

\def\Ls {\ensuremath{\PLambda}\xspace}
\def\Lbar {\ensuremath{\kern 0.1em\overline{\kern -0.1em\PLambda}}\xspace}

\def\Lb      {\ensuremath{\Ls^0_\bquark}\xspace}

\def\BF         {{\ensuremath{\cal B}\xspace}}

\newcommand{\decay}[2]{\ensuremath{#1\!\to #2}\xspace}         

\def\to                 {\ensuremath{\rightarrow}\xspace}

\def\qsq       {\ensuremath{q^2}\xspace}

\newcommand{\etot}{{\ensuremath{\varepsilon_{\rm tot}}}\xspace}

\def\AT#1     {\ensuremath{A_{\mathrm{T}}^{#1}}\xspace}           

\def\C#1      {\ensuremath{\mathcal{C}_{#1}}\xspace}                       
\def\Cp#1     {\ensuremath{\mathcal{C}_{#1}^{'}}\xspace}                    
\def\Ceff#1   {\ensuremath{\mathcal{C}_{#1}^{\mathrm{(eff)}}}\xspace}        
\def\Cpeff#1  {\ensuremath{\mathcal{C}_{#1}^{'\mathrm{(eff)}}}\xspace}       
\def\Ope#1    {\ensuremath{\mathcal{O}_{#1}}\xspace}                       
\def\Opep#1   {\ensuremath{\mathcal{O}_{#1}^{'}}\xspace}                    

\newcommand{\tev}{\ifthenelse{\boolean{inbibliography}}{\ensuremath{~T\kern -0.05em eV}\xspace}{\ensuremath{\mathrm{\,Te\kern -0.1em V}}\xspace}}
\newcommand{\gev}{\ensuremath{\mathrm{\,Ge\kern -0.1em V}}\xspace}
\newcommand{\mev}{\ensuremath{\mathrm{\,Me\kern -0.1em V}}\xspace}
\newcommand{\kev}{\ensuremath{\mathrm{\,ke\kern -0.1em V}}\xspace}
\newcommand{\ev}{\ensuremath{\mathrm{\,e\kern -0.1em V}}\xspace}
\newcommand{\gevc}{\ensuremath{{\mathrm{\,Ge\kern -0.1em V\!/}c}}\xspace}
\newcommand{\mevc}{\ensuremath{{\mathrm{\,Me\kern -0.1em V\!/}c}}\xspace}
\newcommand{\gevcc}{\ensuremath{{\mathrm{\,Ge\kern -0.1em V\!/}c^2}}\xspace}
\newcommand{\gevgevcccc}{\ensuremath{{\mathrm{\,Ge\kern -0.1em V^2\!/}c^4}}\xspace}
\newcommand{\mevcc}{\ensuremath{{\mathrm{\,Me\kern -0.1em V\!/}c^2}}\xspace}

\def\mum  {\ensuremath{\,\upmu\rm m}\xspace}

\def\invfb   {\ensuremath{\mbox{\,fb}^{-1}}\xspace}

\def\ps   {\ensuremath{{\rm \,ps}}\xspace}

\newcommand{\stat}{\ensuremath{\mathrm{\,(stat)}}\xspace}
\newcommand{\syst}{\ensuremath{\mathrm{\,(syst)}}\xspace}

\newcommand{\chisq}{\ensuremath{\chi^2}\xspace}
\newcommand{\chisqndf}{\ensuremath{\chi^2/\mathrm{ndf}}\xspace}
\newcommand{\chisqip}{\ensuremath{\chi^2_{\rm IP}}\xspace}
\newcommand{\chisqvs}{\ensuremath{\chi^2_{\rm VS}}\xspace}
\newcommand{\chisqvtx}{\ensuremath{\chi^2_{\rm vtx}}\xspace}

\def\gsim{{~\raise.15em\hbox{$>$}\kern-.85em
          \lower.35em\hbox{$\sim$}~}\xspace}
\def\lsim{{~\raise.15em\hbox{$<$}\kern-.85em
          \lower.35em\hbox{$\sim$}~}\xspace}

\def\pt         {\mbox{$p_{\rm T}$}\xspace}

\def\mrad{\ensuremath{\rm \,mrad}\xspace}

\def\evtgen     {\mbox{\textsc{EvtGen}}\xspace}

\def\geant      {\mbox{\textsc{Geant4}}\xspace}

\def\photos     {\mbox{\textsc{Photos}}\xspace}

\def\pythia     {\mbox{\textsc{Pythia}}\xspace}

\def\tell1  {TELL1\xspace}
\def\ukl1   {UKL1\xspace}

\newcommand{\eg}{\mbox{\itshape e.g.}\xspace}

%% file: title-LHCb-PAPER.tex

\begin{titlepage}
\pagenumbering{roman}

\vspace*{-1.5cm}
\centerline{\large EUROPEAN ORGANIZATION FOR NUCLEAR RESEARCH (CERN)}
\vspace*{1.5cm}
\hspace*{-0.5cm}
\begin{tabular*}{\linewidth}{lc@{\extracolsep{\fill}}r}
\ifthenelse{\boolean{pdflatex}}
{\vspace*{-2.7cm}\mbox{\!\!\!\includegraphics[width=.14\textwidth]{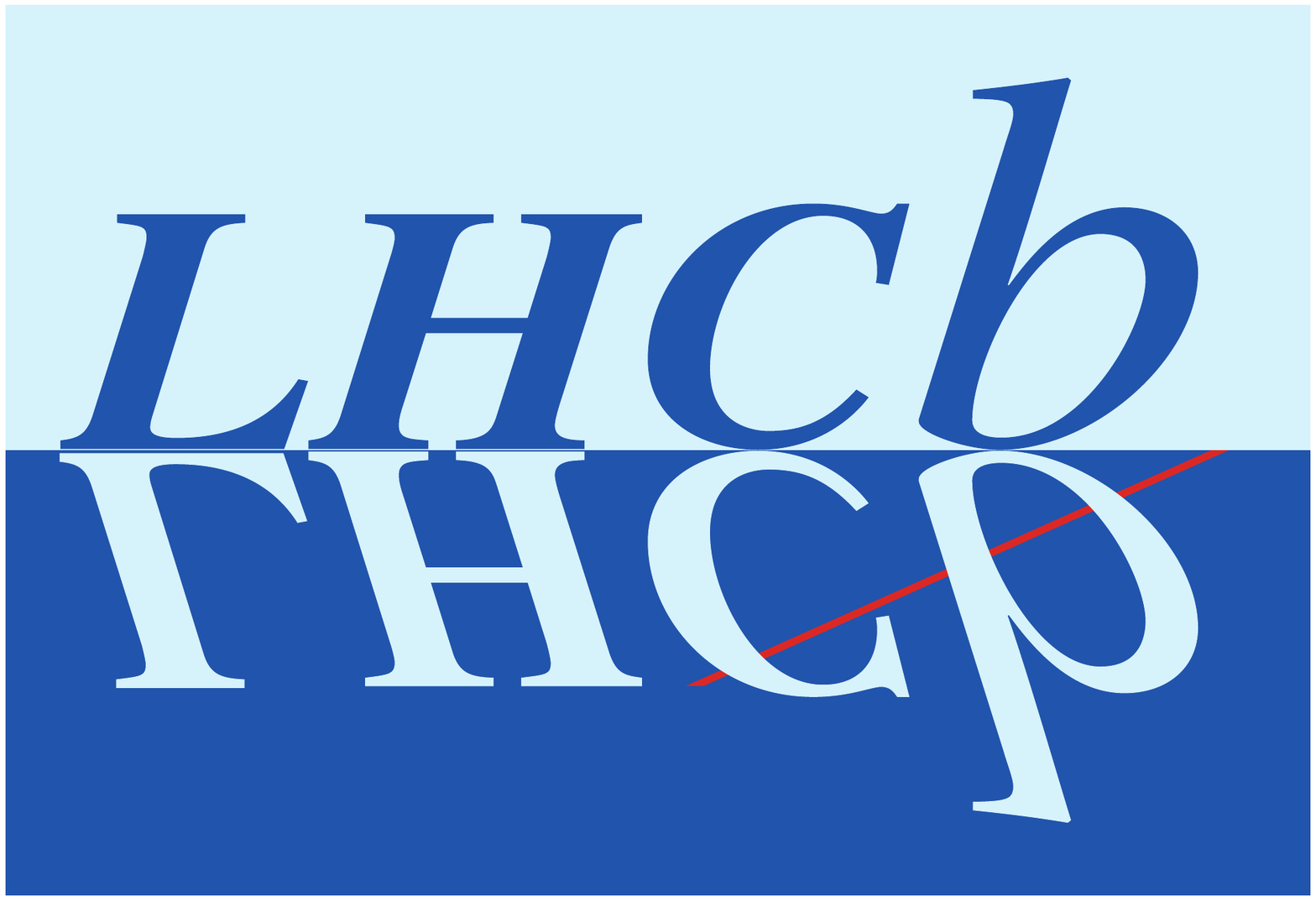}} & &}%
{\vspace*{-1.2cm}\mbox{\!\!\!\includegraphics[width=.12\textwidth]{figs/lhcb-logo.eps}} & &}%
\\
 & & CERN-PH-EP-2013-094 \\  
 & & LHCb-PAPER-2013-025 \\  
 & & 11 June 2013
\end{tabular*}

\vspace*{3.0cm}

{\bf\boldmath\huge
\begin{center}
  Measurement of the differential branching fraction of the decay
  \decay{\Lb}{\Ls\mumu}
\end{center}
}

\vspace*{2.0cm}

\begin{center}
The \lhcb collaboration\footnote{Authors are listed on the following pages.}
\end{center}

\vspace{\fill}

\begin{abstract}
  \noindent
  The differential branching fraction of the decay \decay{\Lb}{\Ls\mumu}
  is measured as a function of  the square of the dimuon
  invariant mass, \qsq.  A yield of $78\pm12$ \decay{\Lb}{\Ls\mumu}
  decays is observed using data, corresponding to an integrated
  luminosity of 1.0\invfb, collected by the \lhcb experiment at a
  centre-of-mass energy of 7\tev. A significant signal is found
  in the \qsq region above the square of the \jpsi mass, while at
  lower-\qsq values upper limits are set on the differential branching
  fraction.  Integrating the differential branching fraction over
  \qsq, while excluding the \jpsi and \psitwos regions, gives a 
  branching fraction of $\BF(\decay{\Lb}{\Ls\mumu})=(0.96\pm
  0.16\stat\pm 0.13\syst\pm 0.21 (\mathrm{norm}))\times 10^{-6}$,
  where the uncertainties are statistical, systematic and due to the
  normalisation mode, \decay{\Lb}{\jpsi\Ls}, respectively.
\end{abstract}

\vspace*{1.0cm}

\begin{center}
  Submitted to Physics Letters B
\end{center}

\vspace{\fill}


{\footnotesize 
\centerline{\copyright~CERN on behalf of the \lhcb collaboration, license \href{http://creativecommons.org/licenses/by/3.0/}{CC-BY-3.0}.}}
\vspace*{2mm}
\end{titlepage}


\newpage
\setcounter{page}{2}
\mbox{~}
\newpage

\input{LHCb_HD_authorlist.tex}

\cleardoublepage

%% file: LHCb_HD_authorlist.tex
\centerline{\large\bf LHCb collaboration}
\begin{flushleft}
\small
R.~Aaij$^{40}$, 
B.~Adeva$^{36}$, 
M.~Adinolfi$^{45}$, 
C.~Adrover$^{6}$, 
A.~Affolder$^{51}$, 
Z.~Ajaltouni$^{5}$, 
J.~Albrecht$^{9}$, 
F.~Alessio$^{37}$, 
M.~Alexander$^{50}$, 
S.~Ali$^{40}$
% ...

{\footnotesize \it
% ...
$ ^{5}$Clermont Universit\'{e}, Universit\'{e} Blaise Pascal, CNRS/IN2P3, LPC, Clermont-Ferrand, France\\
$ ^{6}$CPPM, Aix-Marseille Universit\'{e}, CNRS/IN2P3, Marseille, France\\
% ...
$ ^{9}$Fakult\"{a}t Physik, Technische Universit\"{a}t Dortmund, Dortmund, Germany\\
% ...
$ ^{36}$Universidad de Santiago de Compostela, Santiago de Compostela, Spain\\
$ ^{37}$European Organization for Nuclear Research (CERN), Geneva, Switzerland\\
% ...
$ ^{40}$Nikhef National Institute for Subatomic Physics, Amsterdam, The Netherlands\\
% ...
$ ^{45}$H.H. Wills Physics Laboratory, University of Bristol, Bristol, United Kingdom\\
% ...
$ ^{50}$School of Physics and Astronomy, University of Glasgow, Glasgow, United Kingdom\\
$ ^{51}$Oliver Lodge Laboratory, University of Liverpool, Liverpool, United Kingdom\\
% ...
}
\end{flushleft}

%% file: introduction.tex
\section{Introduction}
The decay \decay{\Lb}{\Ls\mumu} is a rare (\decay{\bquark}{\squark})
flavour-changing neutral current process that in the Standard Model
proceeds through electroweak loop (penguin and \Wpm box) diagrams.
Since non-Standard Model particles may also participate in these loop
diagrams, measurements of this and similar decays can be used to
search for physics beyond the Standard Model. In the past, more
emphasis has been placed on the study of rare decays of mesons than of
baryons, in part due to the theoretical complexity of the latter
\cite{Mannel:1997xy}.  In the particular system studied in this
Letter, the decay products include only a single hadron, simplifying
the theoretical modelling of hadronic physics in the final state.

The study of \Lb baryon decays is of considerable interest for two
reasons.  Firstly, as the \Lb baryon {has non-zero spin}, there
is the potential to improve the limited understanding of the helicity
structure of the underlying Hamiltonian, which cannot be extracted
from mesonic decays \cite{Hiller:2007ur,Mannel:1997xy}. Secondly, as
the composition of the \Lb baryon may be considered as the combination
of a heavy quark with a light diquark system, the hadronic physics
differs significantly from that of the \B meson decay.
This may allow this aspect of the theory to be tested, {which
may lead to improvements in understanding 
of \B mesons}.

Theoretical aspects of the \decay{\Lb}{\Ls\mumu} decay have been
considered both in the SM and in various scenarios of physics beyond
the Standard Model
\cite{Aslam:2008hp,Wang:2008sm,Huang:1998ek,Chen:2001ki,Chen:2001zc,Chen:2001sj,Zolfagharpour:2007eh,Mott:2011cx,Aliev:2010uy,Mohanta:2010eb,Sahoo:2011yb,
Detmold:2012vy,Gutsche:2013pp}.  Although based on the same effective
Hamiltonian as that for the corresponding mesonic transitions, the
hadronic form factors for the \Lb baryon case are less well-known due
to the smaller number of experimental constraints.  This leads to a
large spread in the predicted branching fractions. The differential
branching fraction as a function of the square of the dimuon invariant
mass, $\qsq \equiv m_{\mumu}^2$, is of particular interest.  The
approaches taken by the theoretical calculations depend on the \qsq
region.  By comparing predictions with data as a function of \qsq,
these different methods of treating form factors are tested.
 
The first observation of the decay \decay{\Lb}{\Ls\mumu} by the CDF
collaboration~\cite{Aaltonen:2011qs} had a signal yield of $24\pm5$
events, corresponding to an absolute branching fraction
$\BF(\decay{\Lb}{\Ls\mumu}) =(1.73\pm 0.42\stat\pm 0.55\syst)\times
10^{-6}$, with evidence for signal at \qsq above the
square of the mass of the \psitwos resonance.

Following previous measurements of rare decays involving dimuon final
states \cite{LHCb-PAPER-2012-023,LHCb-PAPER-2013-017}, a first
measurement by \lhcb of the differential and total branching fractions
for the rare decay \decay{\Lb}{\Ls\mumu} is reported. The inclusion of
charge conjugate modes {is implicit throughout}.  The rates are
normalised with respect to the \decay{\Lb}{\jpsi\Ls} decay, with
\decay{\jpsi}{\mumu}.  This analysis uses a $pp$ collision data
sample, corresponding to an integrated luminosity of 1.0\invfb,\
collected during 2011 at a centre-of-mass energy of 7\tev.

%% file: detector.tex
\section{Detector and software}
\label{sec:Detector}
The \lhcb detector~\cite{Alves:2008zz} is a single-arm forward
spectrometer covering the \mbox{pseudorapidity} range $2<\eta <5$,
designed for the study of particles containing \bquark or \cquark
quarks. The detector includes a high-precision tracking system
consisting of a silicon-strip vertex detector (VELO) surrounding the
$pp$ interaction region, a large-area silicon-strip detector located
upstream of a dipole magnet with a bending power of about
$4{\rm\,Tm}$, and three stations of silicon-strip detectors and straw
drift tubes placed downstream. The combined tracking system provides a
momentum measurement with relative uncertainty that varies from
0.4\,\% at 5\gevc to 0.6\,\% at 100\gevc, and impact parameter (IP)
resolution of 20\mum for tracks with high transverse momentum. Charged
hadrons are identified using two ring-imaging Cherenkov
detectors~\cite{LHCb-DP-2012-003}. Photon, electron and hadron
candidates are identified by a calorimeter system consisting of
scintillating-pad and preshower detectors, an electromagnetic
calorimeter and a hadronic calorimeter. Muons are identified by a
system composed of alternating layers of iron and multiwire
proportional chambers~\cite{LHCb-DP-2012-002}.

  The trigger~\cite{LHCb-DP-2012-004} consists of a hardware stage,
based on information from the calorimeter and muon systems, followed
by a software stage, which applies a full event reconstruction.
Candidate events are first required to pass a hardware trigger which
selects muons with a transverse momentum, $\pt>1.48\gevc$. In the
subsequent software trigger, at least one of the final state particles
is required to have both $\pt>0.8\gevc$ and an impact parameter
greater than 100\mum with respect to all of the primary $pp$ interaction
vertices~(PVs) in the event. Finally, the tracks of two or more of the
final state particles are required to form a vertex that is
significantly displaced from the PVs in the event.

A candidate \decay{\Lb}{\Ls\mumu} or \decay{\Lb}{\jpsi\Ls} decay that is
directly responsible for triggering both the {hardware and
software} triggers is denoted as { ``trigger on signal''}.  An
event in which a \Lb baryon is reconstructed in either of these modes
but none of the daughter particles are {necessary for the
trigger decision} is referred to as {``trigger independent of
signal''}.  As these two categories of event are not mutually
exclusive, the overlap may be used to estimate the efficiency of the
trigger {selection} directly from data.

In the simulation, $pp$ collisions are generated using
\pythia~6.4~\cite{Sjostrand:2006za} with a specific \lhcb
configuration~\cite{LHCb-PROC-2010-056}.  Decays of hadronic particles
are described by \evtgen~\cite{Lange:2001uf} in which final state
radiation is generated using \photos~\cite{Golonka:2005pn}. The
interaction of the generated particles with the detector and its
response are implemented using the \geant
toolkit~\cite{Allison:2006ve, *Agostinelli:2002hh} as described
in Ref.~\cite{LHCb-PROC-2011-006}.

%% file: selection.tex
\section{Candidate selection}
 Candidate \decay{\Lb}{\Ls\mumu} (signal mode) and \decay{\Lb}{\jpsi\Ls}
 (normalisation mode) decays are reconstructed from muon, \Ls baryon
 and \jpsi candidates.  The \jpsi candidates are reconstructed via
 their dimuon decays and therefore the \decay{\Lb}{\jpsi\Ls} decay is
 an ideal normalisation process.  The dimuon candidates are formed
 from two oppositely-charged particles identified as muons
 \cite{LHCb-DP-2012-002,LHCb-DP-2012-003}. Good track quality is
 ensured by requiring \chisqndf (\chisq per degree of freedom) $<4$
 for a track fit.  The candidates must also have \chisqip with respect
 to any primary interaction greater than 16, where \chisqip is defined
 as the difference in \chisq of a given PV reconstructed with and
 without the considered track.  These \mumu pairs are required to have
 an invariant mass of less than 5050\mevcc and to be consistent with
 originating from a common vertex ($\chisqvtx/\mathrm{ndf}<9$).

  Candidate \Ls decays are reconstructed in the \decay{\Ls}{\proton\pim}
 mode from two oppositely-charged particles that either both originate
 within the acceptance of the VELO (``long \Ls'' candidates), or both
 originate outside the acceptance of the VELO (``downstream \Ls''
 candidates).  Tracks are required to have $\pt>0.5\gevc$, and \Ls
 candidates must have $\chisqvtx/\mathrm{ndf}<30$ ($<25$ for
 downstream \Ls candidates), a decay time of at least 2\ps, and a
 reconstructed invariant mass within 30\mevcc of the world average
 value \cite{PDG2012}. Due to the distinct kinematics and topology of
 the \Ls decay, it is not necessary to impose particle identification
 requirements on the decay products of the \Ls candidate.

 Candidate $\Lb$\ decays are formed by combining $\Ls$ and dimuon
 candidates that originate from a common vertex
 ($\chisqvtx/\mathrm{ndf}<8$), have $\chisqip<9$, $\chisqvs>100$ and
 an invariant mass in the interval 4.9--7.0\gevcc.  The \chisqvs is
 defined as the difference in \chisq between fits in which the \Lb
 decay vertex is assumed to coincide with the PV and allowing the
 decay vertex to be distinct from the PV.  Candidates must also point
 to the associated PV by requiring the angle between the \Lb momentum
 vector and the vector between the PV and the \Lb decay vertex is less
 than 8\mrad. The associated PV is the one relative to which the \Lb
 candidate has the lowest \chisqip value.

 The final selection is based on a neural network classifier
 \cite{Feindt:2006pm,feindt-2004} with 15 variables as input.  The
 single most important variable is the \chisq from a kinematic fit
 \cite{Hulsbergen:2005pu} that constrains the decay products of the
 \Lb, the \Ls and the dimuon systems to originate from their respective
 vertices.  Other variables that contribute significantly are the
 momentum and transverse momentum of the \Lb\ candidate, the \chisqip
 and track \chisqndf for both muons, the \chisqip of the \Lb\
 candidate, and the separation of the \Ls and \Lb vertices.  Downstream
 and long \Ls decays have separate inputs to the neural network for
 \chisqip and \chisqvs because of the differing track resolution and
 kinematics.  In the final selection of \decay{\Lb}{\jpsi\Ls}
 candidates, the \mumu invariant mass is required to be in the
 interval 3030--3150\mevcc.
 The signal sample used to train the neural network consists of
 simulated \decay{\Lb}{\Ls\mumu} events, while background is taken from
 data in the upper sideband of the \Lb candidate mass spectrum,
 between 6.0 and 7.0\gevcc, which is dominated by candidates with
 dimuon mass in the \jpsi region.  The requirement on the output of
 the neural network is chosen to maximise 
 $N_{\mathrm{S}}/\sqrt{N_{\mathrm{S}}+N_{\mathrm{B}}}$, where
 $N_{\mathrm{S}}$ and $N_{\mathrm{B}}$ are the expected numbers of
 signal and background events, respectively.  To ensure an appropriate
 normalisation of $N_{\mathrm{S}}$, the number of \decay{\Lb}{\jpsi\Ls}
 candidates after the preselection is scaled by the measured ratio of
 branching fractions between the \decay{\Lb}{\Ls\mumu} and
 \decay{\Lb}{\jpsi\Ls} decays~\cite{Aaltonen:2011qs}, and the
 \decay{\jpsi}{\mumu} branching fraction~\cite{PDG2012}.  The value of
 $N_\mathrm{B}$ is derived from the background training sample
 normalised to the number of candidates in the signal region after
 preselection. The \decay{\Lb}{\Ls\mumu} signal candidates exclude the
 \qsq regions of 8.68--10.09\gevgevcccc and 12.86--14.18\gevgevcccc,
 which are dominated by contributions from the \jpsi and \psitwos
 resonances, respectively.  The effect of finite \qsq resolution is
 negligible.  Relative to the preselected event sample, the neural
 network retains $(76.0\pm0.3)\,\%$ of the rare decay signal while
 rejecting $(95.9\pm0.2)\,\%$ of the background.

%% file: physicsBg.tex
\section{Peaking backgrounds}
\label{sec:physbg}
 Backgrounds are studied using simulated samples of \bquark hadrons in
 which the final state includes two muons.  For the
 \decay{\Lb}{\jpsi\Ls} channel, the only significant contribution found
 is from \decay{\Bz}{\jpsi\KS} decays, with \decay{\KS}{\pip\pim},
 which has the same topology as the \decay{\Lb}{\jpsi\Ls} mode. This
 contribution leads to a broad shape that peaks below the \Lb mass
 region and is accommodated in the mass fit described later.

 For the  \decay{\Lb}{\Ls\mumu} channel, sources  of peaking background
 are  considered in the  \qsq ranges  of interest.   The contributions
 identified  are  \decay{\Lb}{\jpsi\Ls} decays  in  which an  energetic
 photon    is   radiated    from    either   of    the   muons,    and
 \decay{\Bz}{\KS\mumu} decays, where  \decay{\KS}{\pip\pim} and a pion
 is  misreconstructed as  a proton.   The  \decay{\Lb}{\jpsi\Ls} decays
 contribute in the \qsq  region just below $m_{\jpsi}^2$, and populate
 a  mass region significantly  below the  \Lb mass.   The contribution
 from  the \decay{\Bz}{\KS\mumu}  decays  is estimated  by taking  the
 number    of    \decay{\Bz}{\jpsi\KS}    events    found    in    the
 \decay{\Lb}{\jpsi\Ls}  fit, and  scaling this  by the  ratio  of world
 average   branching    fractions   between   the    decay   processes
 \decay{\Bz}{\KS\mumu}   and   \decay{\Bz}{\jpsi\KS}  (including   the
 \decay{\jpsi}{\mumu} branching  fraction) \cite{PDG2012}.  This gives
 fewer than 10 events integrated over \qsq, which is small relative to
 the expected total background levels.

%% file: fit.tex
\section{Yields}

\subsection{Fit description}
 The yields of signal and background events in the data are
 determined in the mass range 5.35--5.85\gevcc using unbinned,
 extended maximum likelihood fits, for the \decay{\Lb}{\Ls\mumu} and
 the \decay{\Lb}{\jpsi\Ls} modes.  The likelihood function has the form
\begin{equation}
\mathcal{L}=e^{-(N_\mathrm{S}+N_\mathrm{B}+N_{\mathrm{P}})} \times \prod_{i=1}^{N}[
  N_\mathrm{S}P_{\mathrm{S}}(m_i)+N_\mathrm{B}P_\mathrm{B}(m_i)+N_{\mathrm{P}}P_{\mathrm{P}}(m_i)]
 \;,
\label{eq:uml}
\end{equation}
 where $N_\mathrm{S}$, $N_\mathrm{B}$ and $N_\mathrm{P}$ are number of
 signal, combinatorial and peaking background events, respectively,
 and $P_j(m_i)$ are the corresponding probability density functions
 (PDFs).  The mass of the \Lb candidate, $m_i$, is determined by a
 kinematic fit of the full decay chain in which the proton and pion
 are constrained such that the $\proton\pim$ invariant mass corresponds
 to the \L baryon mass \cite{PDG2012}.

 The signal shape, in both \decay{\Lb}{\Ls\mumu} and
 \decay{\Lb}{\jpsi\Ls} modes, is described by the sum of two Gaussian
 functions that share a common mean but have independent widths.  The
 combinatorial background is parametrised by a first-order 
 polynomial, while the background due to \decay{\Bz}{\jpsi\KS}
 decays is modelled by an exponential function (with
 a cut-off) convolved with a Gaussian function.

 For the \decay{\Lb}{\jpsi\Ls} mode, the widths and common mean in the
 signal parametrisation are free parameters. The contribution of the
 narrower Gaussian function is fixed to be 86\,\% of the total yield
 based on studies with simulated data.  The parameters describing the
 shape of the peaking background are fixed to those derived from
 simulated \decay{\Bz}{\jpsi\KS} decays.

  For the \decay{\Lb}{\Ls\mumu} decay, the signal shape parameters are
  fixed according to the result of the fit to \decay{\Lb}{\jpsi\Ls}
  data.  Studies with simulated data show that the signal shape
  parameters in both decay modes are consistent with one another, the
  only deviations being in the tails of the mass distribution.  These
  are due to small differences in the momentum spectra of the muons
  and energy loss from radiative effects, and are negligible given the
  uncertainties inherent in the size of the current data sample. The
  peaking background is found to be negligible in the \qsq regions
  considered and is therefore excluded from the fit.

\subsection{Fit results}
\begin{figure}
\centering \includegraphics[width=\textwidth]{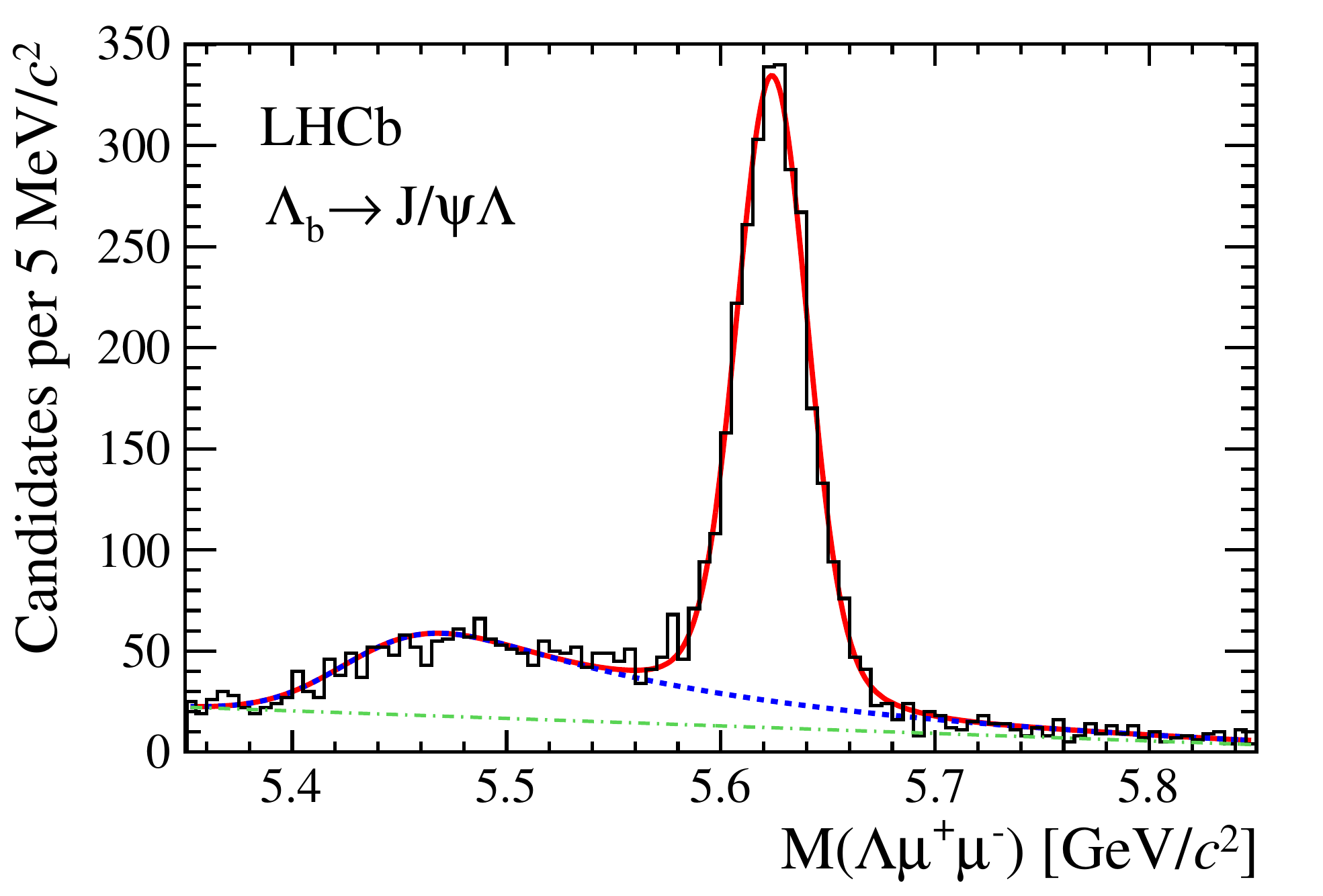}
\caption{\small Invariant mass distribution of the
  \decay{\Lb}{\jpsi\Ls} candidates.  The histogram shows data, the
  solid red line is the overall fit function, the dotted blue line
  represents the sum of the combinatorial and peaking backgrounds and
  the dash-dotted green line the combinatorial background component.}
\label{fig:totalFit}
\end{figure}
 The invariant mass distributions of the \decay{\Lb}{\jpsi\Ls}
 candidates is shown in Fig.~\ref{fig:totalFit}.  The fitted function
 provides a good description of the data, with a \chisq/ndf
 corresponding to a probability of 47\,\%.  The numbers of signal,
 combinatorial background and peaking background events are found to
 be $2680\pm64$, $1294\pm83$ and $1501\pm85$, respectively, and the
 widths of the Gaussian functions are $16.0\pm0.4$ and $33\pm5$\mevcc,
 compatible with simulation.

\begin{figure}
\centering \includegraphics[width=\textwidth]{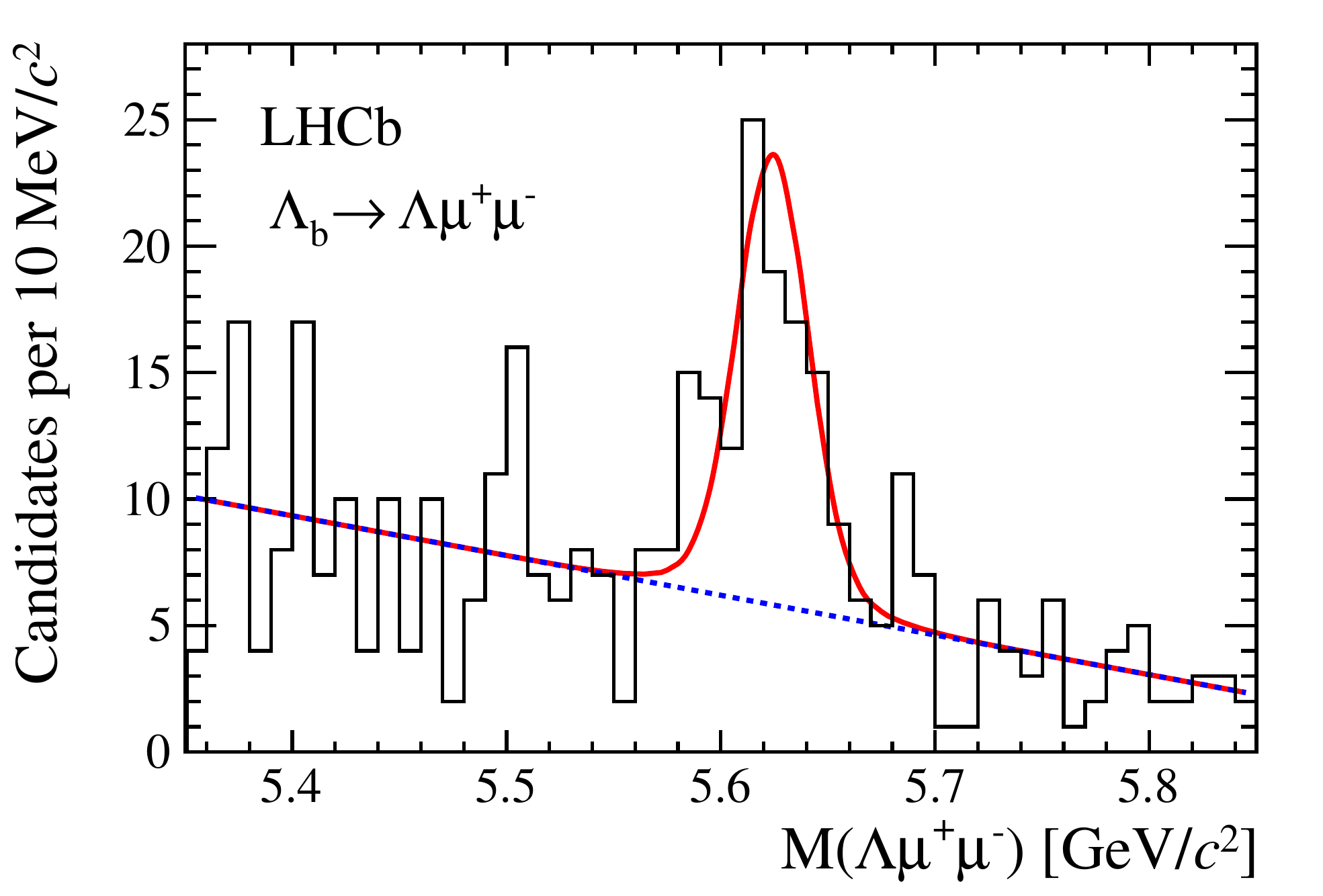}
\caption{\small Invariant mass distribution of the
  \decay{\Lb}{\Ls\mumu} candidates, integrated over all \qsq\ values,
  together with the fit function described in the text.  The histogram
  shows data, the solid red line is the overall fit function
  and the dotted blue line represents the background component.}
\label{fig:totalFitRare}
\end{figure}

\begin{figure}[htb]
\centering \includegraphics[width=\textwidth]{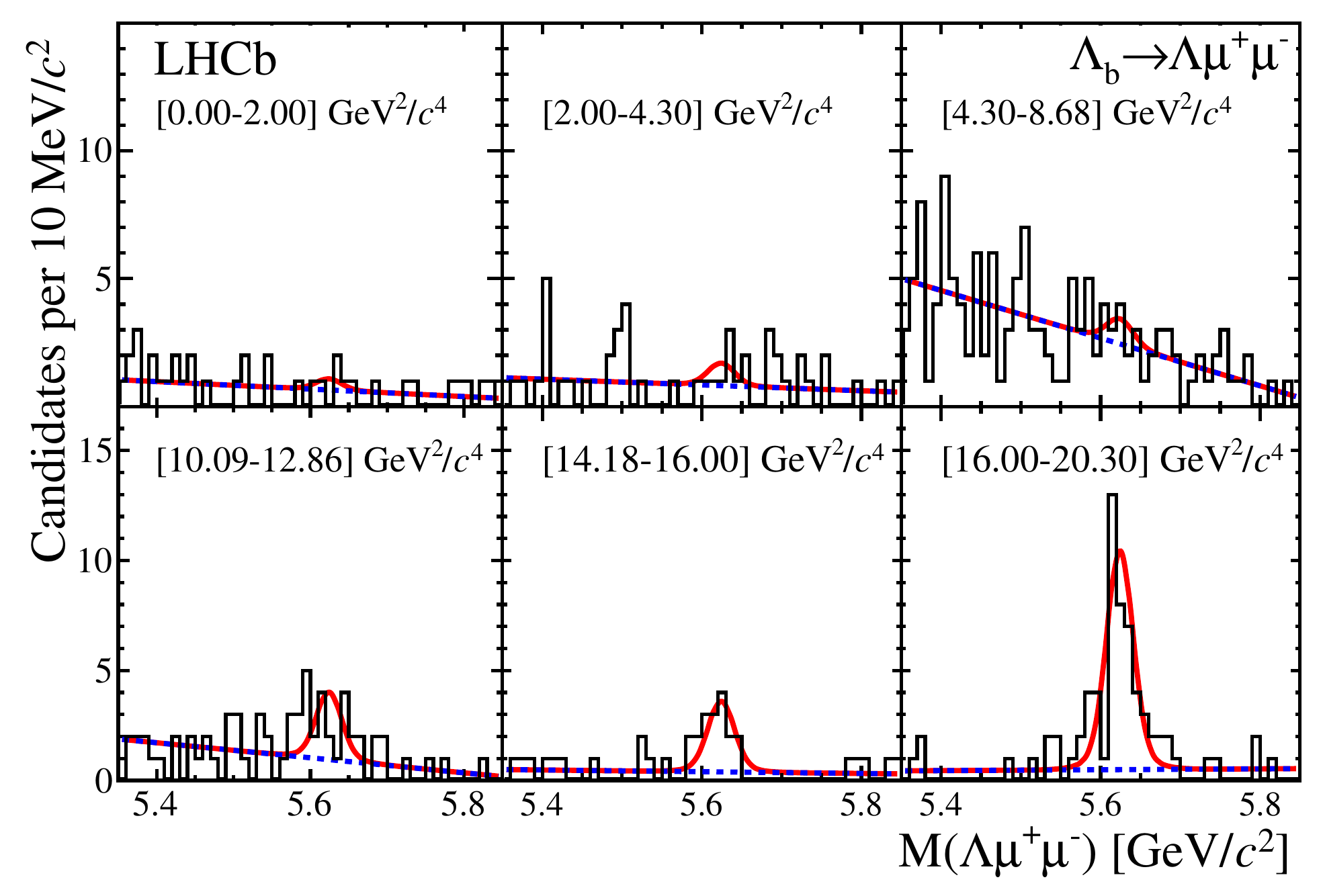}
\caption{\small Invariant mass distributions for the rare decay
  \decay{\Lb}{\Ls\mumu} candidates, in six \qsq\ intervals, together with the
  fit function described in the text. The histogram shows data, the
  solid red line is the overall fit function and the dotted
  blue line represents the background component.}
\label{fig:differentialFit}
\end{figure}

 The invariant mass distribution for the \decay{\Lb}{\Ls\mumu} process,
 integrated over \qsq and in six \qsq intervals, are shown in
 Figs.~\ref{fig:totalFitRare} and \ref{fig:differentialFit},
 respectively. The  yields, both integrated and differential in
 \qsq, are summarised in Table~\ref{tab:rareYields}.  The same \qsq
 intervals as in Ref.~\cite{Aaltonen:2011qs} are used to facilitate
 comparison with the CDF measurements. The statistical significance of
 the observed signal yields in Table~\ref{tab:rareYields} are
 evaluated as $\sqrt{2\Delta\ln{\mathcal{L}}}$, where
 $\Delta\ln{\mathcal{L}}$ is the change in the logarithm of the
 likelihood function when the signal component is excluded from the
 fit, relative to the nominal fit in which it is present.  Significant
 signal yields are only apparent for $\qsq > m_{\jpsi}^2$. Yields
 at lower-\qsq values are compatible with zero, consistent with
 previous observations~\cite{Aaltonen:2011qs}.

\begin{table}
\centering
\caption{\small Signal ($N_\mathrm{S}$) and background
  ($N_\mathrm{B}$) decay yields obtained from the \decay{\Lb}{\Ls\mumu}
  mass fit in each \qsq interval. The integrated yield is the result
  of a fit without separation of the data into distinct \qsq
  regions. The statistical significance is calculated as described in
  the text.}
\label{tab:rareYields}
\begin{tabular}{c|c|c|c}
 $q^2$ interval [\gevgevcccc] & $N_\mathrm{S}$ & $N_\mathrm{B}$ & Significance \\ \hline
  0.00 -- 2.00 & ~$2 \pm 3$~  & $34 \pm 6$  &  0.8\\
  2.00 -- 4.30 & ~$4 \pm 3$~  & $42 \pm 7$  &  1.4\\
 4.30 -- 8.68  & ~$4 \pm 5$~  & $134 \pm 12$&  1.0 \\
 10.09 -- 12.86& $13 \pm 5$~ & $52 \pm 8$ &  3.4 \\
 14.18 -- 16.00& $14 \pm 4$~ & $20 \pm 5$ &  4.9 \\
 16.00 -- 20.30& $44 \pm 7$~ & $24 \pm 6$ &  9.8\\ \hline
 Integrated yield &$\,78 \pm 12$& $310 \pm 19$  &  8.9
\end{tabular}
\end{table}

\clearpage

%% file: efficiency.tex
 \section{Efficiency}
\label{sec:efficiency}

 The measurement of the differential branching fraction of
 \decay{\Lb}{\Ls\mumu} relative to \decay{\Lb}{\jpsi\Ls} benefits from
 the cancellation of several potential sources of systematic
 uncertainty in the ratio of efficiencies, $\varepsilon_{\rm
 rel}=\etot(\decay{\Lb}{\Ls\mumu})/\etot(\decay{\Lb}{\jpsi\Ls})$.
 The efficiency for each of the decays is calculated according to
\begin{equation}
\etot=\varepsilon(\mathrm{geometry})\;\varepsilon(\mathrm{selection}|\mathrm{geometry})\;\varepsilon(\mathrm{trigger}|\mathrm{selection})\;,
\end{equation}
 where the first term represents the efficiency for the final state
 particles to be within the \lhcb angular acceptance, the second term
 the combined efficiency for candidate detection, reconstruction and
 selection, and the rightmost term the efficiency for an event to
 satisfy the trigger requirements if it is reconstructed and selected.
 All efficiencies are evaluated using simulated data.  A phase space
 model is used for \decay{\Lb}{\jpsi\Ls} decays.  The model used for
 \decay{\Lb}{\Ls\mumu} decays includes \qsq and angular dependence as
 described in Ref.~\cite{Aliev:2005np}, together with Wilson
 coefficients based on Refs.~\cite{Buras:1994dj,Buras:1993xp}.
 Interference effects from charmonium contributions are not included.
 
 With these models, the geometric acceptance is found to be 16\,\% for
 \decay{\Lb}{\jpsi\Ls} decays and in the range 16--20\,\% (\qsq
 dependent) for the \decay{\Lb}{\Ls\mumu} channel.
 The overall efficiency to reconstruct and select the
 \decay{\Lb}{\Ls\mumu} decays varies from 1.3\,\% in the lowest \qsq
 interval to values around 2.5\,\% in the higher-\qsq regions.  The
 \decay{\Lb}{\jpsi\Ls} decay has a similar efficiency to the
 larger-\qsq regions of the rare decay.
 The trigger efficiency is calculated using an emulation of the
 hardware trigger, combined with the same software stage of the
 trigger that was used for data.  The trigger efficiency increases
 from approximately 50\,\% to 80\,\% for the lowest to highest \qsq
 regions, respectively.  An independent cross-check of the trigger
 efficiency is performed using \decay{\Lb}{\jpsi\Ls} data by
 calculating the ratio of yields that are both classified as trigger
 on signal and trigger independent of signal relative to those that
 are only classified as trigger independent of signal. This
 data-driven method gives an efficiency of $(75\pm7)$\,\%, which is
 consistent with that of $(70.5\pm0.3)$\,\% computed from simulation.

 The relative efficiency for the ratio of branching fractions in each
 \qsq interval, calculated from the absolute efficiencies described
 above, are given in Table~\ref{tab:relativeTotalEfficiency}.  The
 rise in relative efficiency as a function of increasing \qsq is
 dominated by two effects. Firstly, at low \qsq the muons have lower
 momenta and therefore have a lower probability of satisfying the
 trigger requirements.  Secondly, at low \qsq the \L baryon has a
 larger fraction of the \Lb momentum and is more likely to decay
 outside of the acceptance.  The uncertainties combine both
 statistical and systematic contributions (with the latter dominating)
 and include a small correlated uncertainty due to the use of a single
 sample of \decay{\Lb}{\jpsi\Ls} decays as the normalisation channel
 for all \qsq\ intervals.  The systematic uncertainties are described
 in more detail in Sect.~\ref{sec:systemtics}.

\begin{table}[tbp]
\centering
\caption{\small Total relative efficiency,
$\varepsilon_{\mathrm{rel}}$, between \decay{\Lb}{\Ls\mumu} and
\decay{\Lb}{\jpsi\Ls} decays.  The uncertainties are the combination of
both statistical and systematic components, and are dominated by the
latter.}
\label{tab:relativeTotalEfficiency}
\begin{tabular}{c|c}
\qsq interval [\gevgevcccc] & $\varepsilon_{\mathrm{rel}}$ \\ \hline
0.00--2.00     & $0.48\pm 0.07$ \\
2.00--4.30     & $0.74\pm 0.08$ \\
4.30--8.68     & $0.88\pm 0.09$ \\
10.09--12.86   & $1.19\pm 0.12$ \\
14.18--16.00   & $1.36\pm 0.14$ \\
16.00--20.30   & $1.28\pm 0.15$ \\
\end{tabular}
\end{table}

%% file: systematics.tex
\section{Systematic uncertainties}
\label{sec:systemtics}
\subsection{Yields}
\label{sec:systematics_yields}
 Three separate sources of  systematic uncertainty on the
 measured yields are considered for both the \decay{\Lb}{\jpsi\Ls} and
 \decay{\Lb}{\Ls\mumu} decay modes: the definition of the signal PDF,
 the definition of the background PDF and the choice of the fixed
 parameters used in the fits to data.

 For the \decay{\Lb}{\jpsi\Ls} decays, the default signal PDF is
 replaced by a single Gaussian function.  A 2.0\,\% change in signal
 yield relative to the default fit is observed and assigned as the
 systematic uncertainty.  The shape of the combinatorial background
 function is changed from the default first-order polynomial
 to a second-order  polynomial.  The 1.8\,\% change in the
 signal yield is assigned as the systematic uncertainty.
 To estimate the sensitivity of the background process
 \decay{\Bz}{\jpsi\KS} to differences between data and simulation, the
 shape of this background is varied in the fit. A relative uncertainty
 of 4.7\,\% is assigned.
 For \decay{\Lb}{\Ls\mumu} decays, as the parameter values of the
 signal PDF are from fits to the \decay{\Lb}{\jpsi\Ls} data, the
 uncertainty in the signal shape is accounted for by using the signal
 shape parameters and covariance matrix obtained from the
 \decay{\Lb}{\jpsi\Ls} mass fit. The dependence on the shape of the
 signal PDF is investigated by fitting data using the parameters
 determined from the single-Gaussian function treatment of the
 \decay{\Lb}{\jpsi\Ls} data described above.  The combinatorial
 background modelling is studied in the same way as for the
 \decay{\Lb}{\jpsi\Ls} decays.  The systematic uncertainties on the
 yield in each \qsq interval are summarised in
 Table~\ref{tab:rareYieldSystematics}, where the total is the sum in
 quadrature of the three individual components.
\begin{table}[tbp]
\caption{\small Absolute systematic uncertainties on the yields
for the \decay{\Lb}{\Ls\mumu} decay.}
\label{tab:rareYieldSystematics}
\centering
\begin{tabular}{l|l|l|l|l|l|l}
                  & \multicolumn{6}{c}{\protect\qsq interval [\gevgevcccc] }  \\
Source            & 0.00-- & 2.00-- & 4.30-- & 10.09-- & 14.18-- & 16.00-- \\
                  & 2.00   & 4.30   & 8.68   & 12.86   & 16.00   & 20.30\\ \hline
Signal PDF        &  0.08  & 0.08   & 0.16   &  0.4    & 0.08    &  2.3 \\    
Combinatorial
 background       &  2.7~  & 0.7~   & 0.21   &  3.5    & 2.2~    &  2.5 \\    
Signal shape
parameters        &  0.04  & 0.08   & 0.09   &  0.4    & 0.17    &  1.1 \\ \hline
Total             &  2.7~  & 0.7~   & 0.28   &  3.5    & 2.2~    &  3.5 \\ 

\end{tabular}
\end{table}
 No additional uncertainty is assigned to account for finite peaking 
 background, as constraining it to the prediction from simulated
 \decay{\Bz}{\KS\mumu} decays has a negligible effect.

\subsection{Relative efficiencies}
 In measuring the \qsq dependence of the differential branching
 fraction, three types of correlation are taken into account: those
 between the normalisation and signal decays; those between the
 different \qsq regions; and those between the geometric, selection
 and trigger efficiencies.  For simplicity, correlations among \qsq
 intervals are taken into account where a systematic uncertainty is
 significant and neglected where a given uncertainty is small compared
 to the dominant sources.  Overall, the dominant systematic effect
 identified is that related to the current knowledge of the angular
 structure of the decays and \qsq dependence of the decay channels.
 The uncertainty due to the finite size of simulated samples used is
 comparable to that from other sources considered, and is summarised
 together with all other contributions to the relative efficiency in
 Table~\ref{tab:relativeUnc}, where the total is the sum in
 quadrature of the individual components.

\newlength{\alwidth}
\settowidth{\alwidth}{$<$}
\begin{table}[tbp]
\caption{\small Absolute systematic uncertainties on the total
  relative efficiency, $\varepsilon_{\mathrm{rel}}$.}
\label{tab:relativeUnc}
\centering
\begin{tabular}{l|l|l|l|l|l|l}
                  & \multicolumn{6}{c}{\protect\qsq interval [\gevgevcccc] }  \\
Source            & 0.00--     & 2.00--    & 4.30--  & \makebox[\alwidth]{}10.09-- & 14.18-- & 16.00-- \\
                  & 2.00       &  4.30     &  8.68   & \makebox[\alwidth]{}12.86  &  16.00 &  20.30 \\ \hline
Simulated sample size
                  &  0.014     &  0.015    &  0.015  &  \makebox[\alwidth]{}0.025  &  0.04  &  0.032 \\    
Decay structure   &  0.05      &  0.07     &  0.08   &  \makebox[\alwidth]{}0.11   &  0.13  &  0.12  \\ 
Polarisation      &  0.007     &  0.007    &  0.011   &  \makebox[\alwidth]{}0.014 &  0.015 &  0.05  \\ 
\Ls reconstruction efficiency
                  &  0.027     &  0.009    &  0.003  &  $<$0.001  &  0.003 &  0.004 \\ 
Production kinematics
                  &  0.023     &  0.005    &  0.007   &  \makebox[\alwidth]{}0.026 &  0.014 &  0.05  \\ 
Neural network    &  0.021     &  0.027    &  0.032   &  \makebox[\alwidth]{}0.021 &  0.002&  0.04  \\ \hline
Total             &  0.07~     &  0.08~    &  0.09    &  \makebox[\alwidth]{}0.12   &  0.14  &  0.15  \\ 
\end{tabular}
\end{table}

\subsubsection{Decay structure and production polarisation}
 The main factors that affect the detection efficiencies are the
 angular structure of the decays and the production
 polarisation. Although these arise from different parts of the
 process, the efficiencies are linked and therefore are treated
 together.

 For the \decay{\Lb}{\Ls\mumu} decay, the impact of the limited
 knowledge of the production polarisation, $P_b$, is estimated by
 comparing the default efficiency with that in either of the fully
 polarised scenarios, $P_b=\pm1$, taking the larger difference as the
 associated uncertainty.  To assess the systematic uncertainty due to
 the decay structure, the efficiency from the default model
 \cite{Aliev:2005np,Buras:1994dj,Buras:1993xp} is compared with that
 from the phase space decay, taking the larger of this difference or
 the statistical precision as the systematic uncertainty.

 For the \decay{\Lb}{\jpsi\Ls} mode, the default phase space decay is
 compared with the efficiency derived using the model from
 Ref.~\cite{Hrivnac:1994jx}, which depends on the polarisation
 parameter $P_b$ and four complex amplitudes. While fixing $P_b=0$, a
 scan of the four complex amplitudes is made and the distribution of
 the change in efficiency relative to the default is constructed. The
 sum in quadrature of the mean and r.m.s.\ of this distribution is
 assigned as the systematic uncertainty due to the decay structure.

 To assess the importance of the production polarisation, this
 exercise is repeated while setting $P_b=\pm1$. The sum in quadrature
 of the mean and r.m.s.\ of the distribution of deviations from the
 default gives the combined effect of decay structure and production
 polarisation.  The systematic uncertainty due to production
 polarisation alone is determined by subtracting in quadrature the
 systematic uncertainty due to the decay structure.

 The impact of $P_b$ on the efficiencies is found to be small using
 the fully polarised scenarios, which are a conservative variation
 relative to the recent measurement of Ref.~\cite{LHCb-PAPER-2012-057}.

 \subsubsection{\boldmath Lifetime of \Lb baryon}
 The \Lb baryon lifetime used throughout is 1.425\ps \cite{PDG2012}
 and the systematic uncertainty associated with this assumption is
 investigated by varying the lifetime by one standard deviation
 (0.032\ps).  No significant effect is found.

 \subsubsection{\boldmath Reconstruction efficiency for \Ls baryon}
 The \Ls baryon is reconstructed from either long or downstream tracks,
 and their relative proportions differ between data and simulation.
 For simulated \decay{\Lb}{\jpsi\Ls} decays, $(21.1\pm0.2)$\,\% of \Ls
 baryon candidates are reconstructed from long tracks, compared to
 $(26.4\pm0.7)$\,\% in data.  For the phase space decay distribution
 of simulated \decay{\Lb}{\Ls\mumu} decays, $(21.5\pm0.1)$\,\%
 (integrated over \qsq) are long tracks, indicating that both decay
 modes have a similar behaviour.
 To account for a potential effect due to the different fractions of
 long and downstream tracks observed in data and simulation, the
 efficiencies are first determined separately for \Ls baryon candidates
 formed exclusively from long and from downstream tracks. A new
 relative efficiency is then determined, setting the fraction of
 downstream tracks to 27\,\% for simulated \decay{\Lb}{\jpsi\Ls}
 decays, and increasing it by 5\,\% in each \qsq interval for
 simulated \decay{\Lb}{\Ls\mumu} decays.  The systematic uncertainty
 from this source is assigned as the difference between this
 reweighted efficiency and the default case.

\subsubsection{Production kinematics}
 There is a small difference between data and simulation in the
 momentum and transverse momentum distributions of the \L baryon
 produced in the \decay{\Lb}{\jpsi\Ls} decays. Simulated data are
 reweighted to reproduce these distributions in data, and the
 differences in the relative efficiencies with respect to the default
 are assigned as the systematic uncertainty due to production
 kinematics.

\subsubsection{Modelling of neural network observables}
 A discrepancy is observed between data and simulation in the neural
 network response for \decay{\Lb}{\jpsi\Ls} decay candidates.  This is
 due to differences between $\chi^2$ distributions in data and simulation.
 A systematic uncertainty is assigned as the change relative to the
 default efficiency after all efficiencies are recalculated using
 reweighted neural network input variables.

%% file: summary.tex
\section{Results and conclusion}
 The relative differential branching fraction is measured in each
 \qsq interval as
 \begin{equation}
 \frac{1}{\BF(\decay{\Lb}{\jpsi\Ls})}
 \frac{\mathrm{d}\BF(\decay{\Lb}{\Ls\mumu})}{\mathrm{d}\qsq}=
 \frac{N_\mathrm{S}(\decay{\Lb}{\Ls\mumu})}
 {N_\mathrm{S}(\decay{\Lb}{\jpsi\Ls})}\frac{1}{\varepsilon_{\mathrm{rel}}}\BF(\decay{\jpsi}{\mumu})\frac{1}{\Delta\qsq}\;,
 \label{eq:reldBF}
 \end{equation}
 where $\Delta\qsq$ represents the width of the  given \qsq
 interval.

  For \qsq regions in which no statistically significant signal is
  observed, an upper limit on
  $\mathrm{d}\BF(\decay{\Lb}{\Ls\mumu})/\mathrm{d}\qsq$ is calculated
  using the following Bayesian approach.  The signal PDF for
  \decay{\Lb}{\Ls\mumu} decays is reparametrised in terms of the
  relative differential rate of Eq.~\ref{eq:reldBF},
  $N_\mathrm{S}(\decay{\Lb}{\jpsi\Ls})$, $\varepsilon_{\mathrm{rel}}$ and
  \BF(\decay{\jpsi}{\mumu}).  The known uncertainties on the
  \decay{\Lb}{\jpsi\Ls} yield and $\varepsilon_{\mathrm{rel}}$ are
  included in the fit with Gaussian constraints and the profile
  likelihood over the relative branching fraction is then obtained.  An
  upper limit is set at the value where the posterior likelihood
  corresponds to 90\,\% (95\,\%). A uniform prior between zero and
  $3\times10^{-3}$ is used. The limits on the absolute differential
  branching fractions are given by the product of the relative limit
  and \BF(\decay{\Lb}{\jpsi\Ls}) and include the uncertainty on
  \BF(\decay{\Lb}{\jpsi\Ls}) from Ref.~\cite{PDG2012}.

 \begin{table}[tbp]
 \centering\caption{\small Measured relative differential
 branching fraction,
 $(1/\BF(\decay{\Lb}{\jpsi\Ls}))\;\mathrm{d}\BF(\decay{\Lb}{\Ls\mumu})/\mathrm{d}\qsq$.
 The first uncertainty is statistical and the second is systematic.
 The systematic uncertainty includes the small, correlated component
 due to
 $\BF(\decay{\jpsi}{\mumu})=(5.93\pm0.06)\times10^{-2}$~\cite{PDG2012}. The
 rightmost column gives the 90\,\% (95\,\%) confidence level upper
 limit (UL) on the relative branching fraction in \qsq intervals where
 no significant signal is observed.}
 \protect\label{tab:diffBF}
 \begin{tabular}{c|c|c}
 \multirow{2}{*}{\qsq interval [\gevgevcccc]}  & \multirow{2}{*}{$\dfrac{1}{\BF(\decay{\Lb}{\jpsi\Ls})}
 \dfrac{\mathrm{d}\BF}{\mathrm{d}\qsq}$ $[10^{-4} (\gevgevcccc)^{-1}]$} & \multirow{2}{*}{UL $[10^{-4} (\gevgevcccc)^{-1}]$}\\
   &  &  \\ \hline
 0.00 -- 2.00     &  $ 0.45 \pm 0.62 \pm 0.64 $ & 1.7 (2.1) \\ 
 2.00 -- 4.30     &  $ 0.50 \pm 0.41 \pm 0.11 $ & 1.3 (1.5) \\
 4.30 -- 8.68     &  $ 0.25 \pm 0.27 \pm 0.03 $ & 0.7 (0.9) \\
 10.09 -- 12.86   &  $ 0.90 \pm 0.34 \pm 0.26 $ & -- \\
 14.18 -- 16.00   &  $ 1.26 \pm 0.38 \pm 0.25 $ & -- \\
 16.00 -- 20.30   &  $ 1.76 \pm 0.29 \pm 0.27 $ & -- \\
 \end{tabular}
 \end{table}
 \begin{table}[tbp]
 \centering
 \caption{\small Measured differential branching fraction,
  $\mathrm{d}\BF(\decay{\Lb}{\Ls\mumu})/\mathrm{d}\qsq$, for
  \BF(\decay{\Lb}{\jpsi\Ls})$ =
  (6.2\pm1.4)\times10^{-4}$~\cite{PDG2012}, where the first uncertainty
  is statistical, the second systematic and the third from the
  uncertainty in \BF(\decay{\Lb}{\jpsi\Ls}).  The rightmost column gives
  the 90\,\% (95\,\%) confidence level upper limit (UL) on the
  branching fraction in \qsq intervals where no significant signal is
  observed.}
 \protect\label{tab:diffBF2}
 \begin{tabular}{c|c|c}
 \qsq interval [\gevgevcccc] &
  $\mathrm{d}\BF/\mathrm{d}\qsq$ $[10^{-7} (\gevgevcccc)^{-1}]$
  & UL $[10^{-7} (\gevgevcccc)^{-1}]$ \\ \hline
 0.00 -- 2.00     & $ 0.28 \pm 0.38 \pm 0.40 \pm 0.06 $   & 1.2 (1.5)   \\ 
 2.00 -- 4.30     & $ 0.31 \pm 0.26 \pm 0.07 \pm 0.07 $   & 0.9 (1.1)    \\
 4.30 -- 8.68     & $ 0.15 \pm 0.17 \pm 0.02 \pm 0.03 $   & 0.5 (0.6)    \\
 10.09 -- 12.86   & $ 0.56 \pm 0.21 \pm 0.16 \pm 0.12 $   & --  \\
 14.18 -- 16.00   & $ 0.79 \pm 0.24 \pm 0.15 \pm 0.17 $   & --  \\
 16.00 -- 20.30   & $ 1.10 \pm 0.18 \pm 0.17 \pm 0.24 $   & --  \\
 \end{tabular}
 \end{table}
 \begin{figure}
 \centering
 \includegraphics[width=\textwidth]{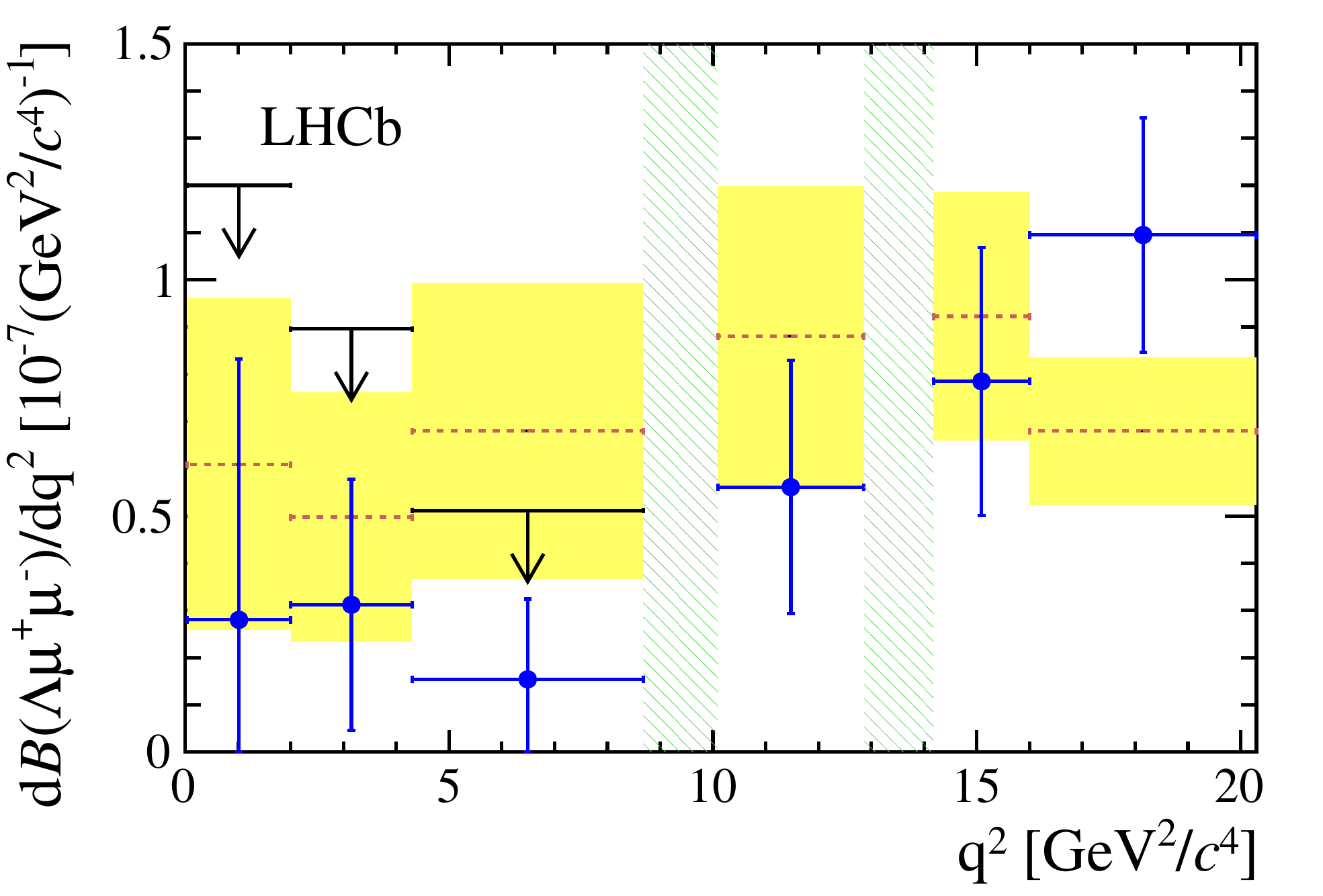}
 \caption{\small Measured differential branching fraction for the
   \decay{\Lb}{\Ls\mumu} decay. In regions without a significant signal,
   the 90\,\% confidence level upper limits are also shown. The
   uncertainties due to components that are fully correlated across all
   \qsq bins, \eg the branching fraction of the normalisation channel
   from Ref.~\cite{PDG2012}, are not included in this figure. The
   dashed red line with the filled area shows the theoretical
   prediction from Ref.~\cite{Detmold:2012vy}.}
 \protect\label{fig:diffBF}
 \end{figure}

  The measured relative differential branching fraction is presented in
  Table~\ref{tab:diffBF}, while the absolute differential branching
  fraction is given in Table~\ref{tab:diffBF2} and shown in
  Fig.~\ref{fig:diffBF}.  The integrated relative branching fraction is
  obtained as the sum of the differential rates in six \qsq intervals
 (weighted by $\Delta\qsq$).  This gives the integral over the full
 phase space, with the exception of the \qsq regions corresponding to
 the \jpsi and \psitwos resonances. In this integration the
 statistical uncertainties are added in quadrature. Systematic
 uncertainties on the \decay{\Lb}{\Ls\mumu} yield and the relative
 efficiency are treated as uncorrelated.  The remaining systematic
 uncertainties, including the statistical and systematic uncertainties
 in the normalisation mode yield from Ref.~\cite{PDG2012}, are treated
 as fully correlated. This leads to the relative branching fraction of
\begin{equation*}
\frac{\BF(\decay{\Lb}{\Ls\mumu})}{\BF(\decay{\Lb}{\jpsi\Ls})}=
(1.54\pm 0.30\stat\pm 0.20\syst\pm 0.02\,(\mathrm{norm}))\times
10^{-3}\;,
\end{equation*}
which corresponds to the absolute branching fraction
\begin{equation*}
\BF(\decay{\Lb}{\Ls\mumu})=(0.96\pm 0.16\stat\pm 0.13\syst\pm 0.21\,(\mathrm{norm}))\times 10^{-6}\; ,
\end{equation*}
 where the last uncertainty accounts for the branching
 fraction of the normalisation mode~\cite{PDG2012}.

 These new measurements of the branching fraction and differential
 branching fraction for the rare decay \decay{\Lb}{\Ls\mumu} are based
 on a yield of $78\pm12$ signal decays obtained from  data, corresponding to an integrated
 luminosity of 1.0\invfb, collected at a centre-of-mass energy of
 7\tev.  Evidence for this process is found for
 $\qsq>m_{\jpsi}^2$ and is compatible with previous measurements by
 the CDF collaboration \cite{Aaltonen:2011qs}.
 Within the precision of measurements presented in this Letter, the
 Standard Model predictions of Ref.~\cite{Detmold:2012vy} provide a
 good description of the data.
\newpage

%% file: acknowledgements.tex
\section*{Acknowledgements}

\noindent We express our gratitude to our colleagues in the CERN
accelerator departments for the excellent performance of the LHC. We
thank the technical and administrative staff at the LHCb
institutes. We acknowledge support from CERN and from the national
agencies: CAPES, CNPq, FAPERJ and FINEP (Brazil); NSFC (China);
CNRS/IN2P3 and Region Auvergne (France); BMBF, DFG, HGF and MPG
(Germany); SFI (Ireland); INFN (Italy); FOM and NWO (The Netherlands);
SCSR (Poland); ANCS/IFA (Romania); MinES, Rosatom, RFBR and NRC
``Kurchatov Institute'' (Russia); MinECo, XuntaGal and GENCAT (Spain);
SNSF and SER (Switzerland); NAS Ukraine (Ukraine); STFC (United
Kingdom); NSF (USA). We also acknowledge the support received from the
ERC under FP7. The Tier1 computing centres are supported by IN2P3
(France), KIT and BMBF (Germany), INFN (Italy), NWO and SURF (The
Netherlands), PIC (Spain), GridPP (United Kingdom). We are thankful
for the computing resources put at our disposal by Yandex LLC
(Russia), as well as to the communities behind the multiple open
source software packages that we depend on.

%% file: main.bbl
\ifx\mcitethebibliography\mciteundefinedmacro
\PackageError{LHCb.bst}{mciteplus.sty has not been loaded}
{This bibstyle requires the use of the mciteplus package.}\fi
\providecommand{\href}[2]{#2}
\begin{mcitethebibliography}{10}
\mciteSetBstSublistMode{n}
\mciteSetBstMaxWidthForm{subitem}{\alph{mcitesubitemcount})}
\mciteSetBstSublistLabelBeginEnd{\mcitemaxwidthsubitemform\space}
{\relax}{\relax}

\bibitem{Mannel:1997xy}
T.~Mannel and S.~Recksiegel, \ifthenelse{\boolean{articletitles}}{{\it
  {Flavor-changing neutral current decays of heavy baryons. The case $\Lb\to\Ls
  \gamma$}}, }{}\href{http://dx.doi.org/10.1088/0954-3899/24/5/006}{J.\ Phys.\
  {\bf G24} (1998) 979}, \href{http://arxiv.org/abs/hep-ph/9701399}{{\tt
  arXiv:hep-ph/9701399}}\relax
\mciteBstWouldAddEndPuncttrue
\mciteSetBstMidEndSepPunct{\mcitedefaultmidpunct}
{\mcitedefaultendpunct}{\mcitedefaultseppunct}\relax
\EndOfBibitem
\bibitem{Hiller:2007ur}
G.~Hiller, M.~Knecht, F.~Legger, and T.~Schietinger,
  \ifthenelse{\boolean{articletitles}}{{\it {Photon polarization from helicity
  suppression in radiative decays of polarized \Lb to spin-3/2 baryons}},
  }{}\href{http://dx.doi.org/10.1016/j.physletb.2007.03.056}{Phys.\ Lett.\
  {\bf B649} (2007) 152}, \href{http://arxiv.org/abs/hep-ph/0702191}{{\tt
  arXiv:hep-ph/0702191}}\relax
\mciteBstWouldAddEndPuncttrue
\mciteSetBstMidEndSepPunct{\mcitedefaultmidpunct}
{\mcitedefaultendpunct}{\mcitedefaultseppunct}\relax
\EndOfBibitem
\bibitem{Aslam:2008hp}
M.~J. Aslam, Y.-M. Wang, and C.-D. Lu,
  \ifthenelse{\boolean{articletitles}}{{\it {Exclusive semileptonic decays of
  $\Lb\to\Ls\ellp\ellm$ in supersymmetric theories}},
  }{}\href{http://dx.doi.org/10.1103/PhysRevD.78.114032}{Phys.\ Rev.\  {\bf
  D78} (2008) 114032}, \href{http://arxiv.org/abs/0808.2113}{{\tt
  arXiv:0808.2113}}\relax
\mciteBstWouldAddEndPuncttrue
\mciteSetBstMidEndSepPunct{\mcitedefaultmidpunct}
{\mcitedefaultendpunct}{\mcitedefaultseppunct}\relax
\EndOfBibitem
\bibitem{Wang:2008sm}
Y.-M. Wang, Y.~Li, and C.-D. Lu, \ifthenelse{\boolean{articletitles}}{{\it
  {Rare decays of $\Lb\to\Ls+\gamma$ and $\Lb\to\Ls +\ellp\ellm$ in the
  light-cone sum rules}},
  }{}\href{http://dx.doi.org/10.1140/epjc/s10052-008-0846-5}{Eur.\ Phys.\ J.\
  {\bf C59} (2009) 861}, \href{http://arxiv.org/abs/0804.0648}{{\tt
  arXiv:0804.0648}}\relax
\mciteBstWouldAddEndPuncttrue
\mciteSetBstMidEndSepPunct{\mcitedefaultmidpunct}
{\mcitedefaultendpunct}{\mcitedefaultseppunct}\relax
\EndOfBibitem
\bibitem{Huang:1998ek}
C.-S. Huang and H.-G. Yan, \ifthenelse{\boolean{articletitles}}{{\it {Exclusive
  rare decays of heavy baryons to light baryons: $\Lb\to\Ls\gamma$ and
  $\Lb\to\Ls \ellp\ellm$ }},
  }{}\href{http://dx.doi.org/10.1103/PhysRevD.59.114022,
  10.1103/PhysRevD.61.039901}{Phys.\ Rev.\  {\bf D59} (1999) 114022},
  \href{http://arxiv.org/abs/hep-ph/9811303}{{\tt arXiv:hep-ph/9811303}}\relax
\mciteBstWouldAddEndPuncttrue
\mciteSetBstMidEndSepPunct{\mcitedefaultmidpunct}
{\mcitedefaultendpunct}{\mcitedefaultseppunct}\relax
\EndOfBibitem
\bibitem{Chen:2001ki}
C.-H. Chen and C.~Q. Geng, \ifthenelse{\boolean{articletitles}}{{\it {Rare
  $\Lb\to\Ls \ellp\ellm$ decays with polarized \L}},
  }{}\href{http://dx.doi.org/10.1103/PhysRevD.63.114024}{Phys.\ Rev.\  {\bf
  D63} (2001) 114024}, \href{http://arxiv.org/abs/hep-ph/0101171}{{\tt
  arXiv:hep-ph/0101171}}\relax
\mciteBstWouldAddEndPuncttrue
\mciteSetBstMidEndSepPunct{\mcitedefaultmidpunct}
{\mcitedefaultendpunct}{\mcitedefaultseppunct}\relax
\EndOfBibitem
\bibitem{Chen:2001zc}
C.-H. Chen and C.~Q. Geng, \ifthenelse{\boolean{articletitles}}{{\it {Baryonic
  rare decays of $\Lb\to\Ls \ellp\ellm$}},
  }{}\href{http://dx.doi.org/10.1103/PhysRevD.64.074001}{Phys.\ Rev.\  {\bf
  D64} (2001) 074001}, \href{http://arxiv.org/abs/hep-ph/0106193}{{\tt
  arXiv:hep-ph/0106193}}\relax
\mciteBstWouldAddEndPuncttrue
\mciteSetBstMidEndSepPunct{\mcitedefaultmidpunct}
{\mcitedefaultendpunct}{\mcitedefaultseppunct}\relax
\EndOfBibitem
\bibitem{Chen:2001sj}
C.-H. Chen and C.~Q. Geng, \ifthenelse{\boolean{articletitles}}{{\it {Lepton
  asymmetries in heavy baryon decays of $\Lb\to\Ls \ellp\ellm$}},
  }{}\href{http://dx.doi.org/10.1016/S0370-2693(01)00937-6}{Phys.\ Lett.\  {\bf
  B516} (2001) 327}, \href{http://arxiv.org/abs/hep-ph/0101201}{{\tt
  arXiv:hep-ph/0101201}}\relax
\mciteBstWouldAddEndPuncttrue
\mciteSetBstMidEndSepPunct{\mcitedefaultmidpunct}
{\mcitedefaultendpunct}{\mcitedefaultseppunct}\relax
\EndOfBibitem
\bibitem{Zolfagharpour:2007eh}
F.~Zolfagharpour and V.~Bashiry, \ifthenelse{\boolean{articletitles}}{{\it
  {Double lepton polarization in $\Lb\to\Ls \ellp\ellm$ decay in the Standard
  Model with fourth generations scenario}},
  }{}\href{http://dx.doi.org/10.1016/j.nuclphysb.2007.12.022}{Nucl.\ Phys.\
  {\bf B796} (2008) 294}, \href{http://arxiv.org/abs/0707.4337}{{\tt
  arXiv:0707.4337}}\relax
\mciteBstWouldAddEndPuncttrue
\mciteSetBstMidEndSepPunct{\mcitedefaultmidpunct}
{\mcitedefaultendpunct}{\mcitedefaultseppunct}\relax
\EndOfBibitem
\bibitem{Mott:2011cx}
L.~Mott and W.~Roberts, \ifthenelse{\boolean{articletitles}}{{\it {Rare
  dileptonic decays of \Lb in a quark model}},
  }{}\href{http://dx.doi.org/10.1142/S0217751X12500169}{Int.\ J.\ Mod.\ Phys.\
  {\bf A27} (2012) 1250016}, \href{http://arxiv.org/abs/1108.6129}{{\tt
  arXiv:1108.6129}}\relax
\mciteBstWouldAddEndPuncttrue
\mciteSetBstMidEndSepPunct{\mcitedefaultmidpunct}
{\mcitedefaultendpunct}{\mcitedefaultseppunct}\relax
\EndOfBibitem
\bibitem{Aliev:2010uy}
T.~M. Aliev, K.~Azizi, and M.~Savci, \ifthenelse{\boolean{articletitles}}{{\it
  {Analysis of the $\Lb\to\Ls \ellp\ellm$ decay in QCD}},
  }{}\href{http://dx.doi.org/10.1103/PhysRevD.81.056006}{Phys.\ Rev.\  {\bf
  D81} (2010) 056006}, \href{http://arxiv.org/abs/1001.0227}{{\tt
  arXiv:1001.0227}}\relax
\mciteBstWouldAddEndPuncttrue
\mciteSetBstMidEndSepPunct{\mcitedefaultmidpunct}
{\mcitedefaultendpunct}{\mcitedefaultseppunct}\relax
\EndOfBibitem
\bibitem{Mohanta:2010eb}
R.~Mohanta and A.~K. Giri, \ifthenelse{\boolean{articletitles}}{{\it {Fourth
  generation effect on $\Lb$ decays}},
  }{}\href{http://dx.doi.org/10.1103/PhysRevD.82.094022}{Phys.\ Rev.\  {\bf
  D82} (2010) 094022}, \href{http://arxiv.org/abs/1010.1152}{{\tt
  arXiv:1010.1152}}\relax
\mciteBstWouldAddEndPuncttrue
\mciteSetBstMidEndSepPunct{\mcitedefaultmidpunct}
{\mcitedefaultendpunct}{\mcitedefaultseppunct}\relax
\EndOfBibitem
\bibitem{Sahoo:2011yb}
S.~Sahoo, C.~K. Das, and L.~Maharana, \ifthenelse{\boolean{articletitles}}{{\it
  {Effect of both Z and Z'-mediated flavor-changing neutral currents on the
  baryonic rare decay $\Lb\to\Ls \ellp\ellm $}},
  }{}\href{http://dx.doi.org/10.1142/S0217751X09047727}{Int.\ ~J.\ ~Mod.\
  ~Phys.\  {\bf A24} (2009) 6223}, \href{http://arxiv.org/abs/1112.4563}{{\tt
  arXiv:1112.4563}}\relax
\mciteBstWouldAddEndPuncttrue
\mciteSetBstMidEndSepPunct{\mcitedefaultmidpunct}
{\mcitedefaultendpunct}{\mcitedefaultseppunct}\relax
\EndOfBibitem
\bibitem{Detmold:2012vy}
W.~Detmold, C.-J.~D. Lin, S.~Meinel, and M.~Wingate,
  \ifthenelse{\boolean{articletitles}}{{\it {$\Lb\to\Ls \ellp\ellm$ form
  factors and differential branching fraction from lattice QCD}},
  }{}\href{http://dx.doi.org/10.1103/PhysRevD.87.074502}{Phys.\ Rev.\  {\bf
  D87} (2013) 074502}, \href{http://arxiv.org/abs/1212.4827}{{\tt
  arXiv:1212.4827}}\relax
\mciteBstWouldAddEndPuncttrue
\mciteSetBstMidEndSepPunct{\mcitedefaultmidpunct}
{\mcitedefaultendpunct}{\mcitedefaultseppunct}\relax
\EndOfBibitem
\bibitem{Gutsche:2013pp}
T.~Gutsche {\em et~al.}, \ifthenelse{\boolean{articletitles}}{{\it {Rare baryon
  decays $\Lb \rightarrow \Ls \ellp \ellm$ $(\ell=e,$ $\mu$, $\tau)$ and $\Lb
  \rightarrow \Ls \gamma$: differential and total rates, lepton- and
  hadron-side forward-backward asymmetries}},
  }{}\href{http://dx.doi.org/10.1103/PhysRevD.87.074031}{Phys.\ Rev.\  {\bf
  D87} (2013) 074031}, \href{http://arxiv.org/abs/1301.3737}{{\tt
  arXiv:1301.3737}}\relax
\mciteBstWouldAddEndPuncttrue
\mciteSetBstMidEndSepPunct{\mcitedefaultmidpunct}
{\mcitedefaultendpunct}{\mcitedefaultseppunct}\relax
\EndOfBibitem
\bibitem{Aaltonen:2011qs}
CDF collaboration, T.~Aaltonen {\em et~al.},
  \ifthenelse{\boolean{articletitles}}{{\it {Observation of the baryonic
  flavor-changing neutral current decay $\Lb \to \Ls \mumu$}},
  }{}\href{http://dx.doi.org/10.1103/PhysRevLett.107.201802}{Phys.\ ~Rev.\
  ~Lett.\  {\bf 107} (2011) 201802}, \href{http://arxiv.org/abs/1107.3753}{{\tt
  arXiv:1107.3753}}\relax
\mciteBstWouldAddEndPuncttrue
\mciteSetBstMidEndSepPunct{\mcitedefaultmidpunct}
{\mcitedefaultendpunct}{\mcitedefaultseppunct}\relax
\EndOfBibitem
\bibitem{LHCb-PAPER-2012-023}
LHCb collaboration, R.~Aaij {\em et~al.},
  \ifthenelse{\boolean{articletitles}}{{\it {Search for the rare decay $\KS \to
  \mu^+ \mu^-$}}, }{}\href{http://dx.doi.org/10.1007/JHEP01(2013)090}{JHEP {\bf
  01} (2013) 90}, \href{http://arxiv.org/abs/1209.4029}{{\tt
  arXiv:1209.4029}}\relax
\mciteBstWouldAddEndPuncttrue
\mciteSetBstMidEndSepPunct{\mcitedefaultmidpunct}
{\mcitedefaultendpunct}{\mcitedefaultseppunct}\relax
\EndOfBibitem
\bibitem{LHCb-PAPER-2013-017}
LHCb collaboration, R.~Aaij {\em et~al.},
  \ifthenelse{\boolean{articletitles}}{{\it {Differential branching fraction
  and angular analysis of the decay $B_{s}^{0}\rightarrow\phi\mu^+\mu^-$}},
  }{}\href{http://arxiv.org/abs/1305.2168}{{\tt arXiv:1305.2168}}, {submitted
  to JHEP}\relax
\mciteBstWouldAddEndPuncttrue
\mciteSetBstMidEndSepPunct{\mcitedefaultmidpunct}
{\mcitedefaultendpunct}{\mcitedefaultseppunct}\relax
\EndOfBibitem
\bibitem{Alves:2008zz}
LHCb collaboration, A.~A. Alves~Jr. {\em et~al.},
  \ifthenelse{\boolean{articletitles}}{{\it {The \lhcb detector at the LHC}},
  }{}\href{http://dx.doi.org/10.1088/1748-0221/3/08/S08005}{JINST {\bf 3}
  (2008) S08005}\relax
\mciteBstWouldAddEndPuncttrue
\mciteSetBstMidEndSepPunct{\mcitedefaultmidpunct}
{\mcitedefaultendpunct}{\mcitedefaultseppunct}\relax
\EndOfBibitem
\bibitem{LHCb-DP-2012-003}
M.~Adinolfi {\em et~al.}, \ifthenelse{\boolean{articletitles}}{{\it
  {Performance of the \lhcb RICH detector at the LHC}},
  }{}\href{http://dx.doi.org/10.1140/epjc/s10052-013-2431-9}{Eur.\ Phys.\ J.\
  {\bf C73} (2013) 2431}, \href{http://arxiv.org/abs/1211.6759}{{\tt
  arXiv:1211.6759}}\relax
\mciteBstWouldAddEndPuncttrue
\mciteSetBstMidEndSepPunct{\mcitedefaultmidpunct}
{\mcitedefaultendpunct}{\mcitedefaultseppunct}\relax
\EndOfBibitem
\bibitem{LHCb-DP-2012-002}
A.~A. Alves~Jr {\em et~al.}, \ifthenelse{\boolean{articletitles}}{{\it
  {Performance of the LHCb muon system}},
  }{}\href{http://dx.doi.org/10.1088/1748-0221/8/02/P02022}{JINST {\bf 8}
  (2013) P02022}, \href{http://arxiv.org/abs/1211.1346}{{\tt
  arXiv:1211.1346}}\relax
\mciteBstWouldAddEndPuncttrue
\mciteSetBstMidEndSepPunct{\mcitedefaultmidpunct}
{\mcitedefaultendpunct}{\mcitedefaultseppunct}\relax
\EndOfBibitem
\bibitem{LHCb-DP-2012-004}
R.~Aaij {\em et~al.}, \ifthenelse{\boolean{articletitles}}{{\it {The \lhcb
  trigger and its performance in 2011}},
  }{}\href{http://dx.doi.org/10.1088/1748-0221/8/04/P04022}{JINST {\bf 8}
  (2013) P04022}, \href{http://arxiv.org/abs/1211.3055}{{\tt
  arXiv:1211.3055}}\relax
\mciteBstWouldAddEndPuncttrue
\mciteSetBstMidEndSepPunct{\mcitedefaultmidpunct}
{\mcitedefaultendpunct}{\mcitedefaultseppunct}\relax
\EndOfBibitem
\bibitem{Sjostrand:2006za}
T.~Sj\"{o}strand, S.~Mrenna, and P.~Skands,
  \ifthenelse{\boolean{articletitles}}{{\it {PYTHIA 6.4 physics and manual}},
  }{}\href{http://dx.doi.org/10.1088/1126-6708/2006/05/026}{JHEP {\bf 05}
  (2006) 026}, \href{http://arxiv.org/abs/hep-ph/0603175}{{\tt
  arXiv:hep-ph/0603175}}\relax
\mciteBstWouldAddEndPuncttrue
\mciteSetBstMidEndSepPunct{\mcitedefaultmidpunct}
{\mcitedefaultendpunct}{\mcitedefaultseppunct}\relax
\EndOfBibitem
\bibitem{LHCb-PROC-2010-056}
I.~Belyaev {\em et~al.}, \ifthenelse{\boolean{articletitles}}{{\it {Handling of
  the generation of primary events in \gauss, the \lhcb simulation framework}},
  }{}\href{http://dx.doi.org/10.1109/NSSMIC.2010.5873949}{Nuclear Science
  Symposium Conference Record (NSS/MIC) {\bf IEEE} (2010) 1155}\relax
\mciteBstWouldAddEndPuncttrue
\mciteSetBstMidEndSepPunct{\mcitedefaultmidpunct}
{\mcitedefaultendpunct}{\mcitedefaultseppunct}\relax
\EndOfBibitem
\bibitem{Lange:2001uf}
D.~J. Lange, \ifthenelse{\boolean{articletitles}}{{\it {The EvtGen particle
  decay simulation package}},
  }{}\href{http://dx.doi.org/10.1016/S0168-9002(01)00089-4}{Nucl.\ Instrum.\
  Meth.\  {\bf A462} (2001) 152}\relax
\mciteBstWouldAddEndPuncttrue
\mciteSetBstMidEndSepPunct{\mcitedefaultmidpunct}
{\mcitedefaultendpunct}{\mcitedefaultseppunct}\relax
\EndOfBibitem
\bibitem{Golonka:2005pn}
P.~Golonka and Z.~Was, \ifthenelse{\boolean{articletitles}}{{\it {PHOTOS Monte
  Carlo: a precision tool for QED corrections in $Z$ and $W$ decays}},
  }{}\href{http://dx.doi.org/10.1140/epjc/s2005-02396-4}{Eur.\ Phys.\ J.\  {\bf
  C45} (2006) 97}, \href{http://arxiv.org/abs/hep-ph/0506026}{{\tt
  arXiv:hep-ph/0506026}}\relax
\mciteBstWouldAddEndPuncttrue
\mciteSetBstMidEndSepPunct{\mcitedefaultmidpunct}
{\mcitedefaultendpunct}{\mcitedefaultseppunct}\relax
\EndOfBibitem
\bibitem{Allison:2006ve}
GEANT4 collaboration, J.~Allison {\em et~al.},
  \ifthenelse{\boolean{articletitles}}{{\it {Geant4 developments and
  applications}}, }{}\href{http://dx.doi.org/10.1109/TNS.2006.869826}{IEEE
  Trans.\ Nucl.\ Sci.\  {\bf 53} (2006) 270}\relax
\mciteBstWouldAddEndPuncttrue
\mciteSetBstMidEndSepPunct{\mcitedefaultmidpunct}
{\mcitedefaultendpunct}{\mcitedefaultseppunct}\relax
\EndOfBibitem
\bibitem{Agostinelli:2002hh}
GEANT4 collaboration, S.~Agostinelli {\em et~al.},
  \ifthenelse{\boolean{articletitles}}{{\it {GEANT4: A simulation toolkit}},
  }{}\href{http://dx.doi.org/10.1016/S0168-9002(03)01368-8}{Nucl.\ Instrum.\
  Meth.\  {\bf A506} (2003) 250}\relax
\mciteBstWouldAddEndPuncttrue
\mciteSetBstMidEndSepPunct{\mcitedefaultmidpunct}
{\mcitedefaultendpunct}{\mcitedefaultseppunct}\relax
\EndOfBibitem
\bibitem{LHCb-PROC-2011-006}
M.~Clemencic {\em et~al.}, \ifthenelse{\boolean{articletitles}}{{\it {The \lhcb
  simulation application, \gauss: design, evolution and experience}},
  }{}\href{http://dx.doi.org/10.1088/1742-6596/331/3/032023}{{J.\ of Phys.\ :
  Conf.\ Ser.\ } {\bf 331} (2011) 032023}\relax
\mciteBstWouldAddEndPuncttrue
\mciteSetBstMidEndSepPunct{\mcitedefaultmidpunct}
{\mcitedefaultendpunct}{\mcitedefaultseppunct}\relax
\EndOfBibitem
\bibitem{PDG2012}
Particle Data Group, J.~Beringer {\em et~al.},
  \ifthenelse{\boolean{articletitles}}{{\it {\href{http://pdg.lbl.gov/}{Review
  of particle physics}}}, }{}Phys.\ Rev.\  {\bf D86} (2012) 010001\relax
\mciteBstWouldAddEndPuncttrue
\mciteSetBstMidEndSepPunct{\mcitedefaultmidpunct}
{\mcitedefaultendpunct}{\mcitedefaultseppunct}\relax
\EndOfBibitem
\bibitem{Feindt:2006pm}
M.~Feindt and U.~Kerzel, \ifthenelse{\boolean{articletitles}}{{\it {The
  NeuroBayes neural network package}},
  }{}\href{http://dx.doi.org/10.1016/j.nima.2005.11.166}{Nucl.\ Instrum.\
  Meth.\  {\bf A559} (2006) 190}\relax
\mciteBstWouldAddEndPuncttrue
\mciteSetBstMidEndSepPunct{\mcitedefaultmidpunct}
{\mcitedefaultendpunct}{\mcitedefaultseppunct}\relax
\EndOfBibitem
\bibitem{feindt-2004}
M.~Feindt, \ifthenelse{\boolean{articletitles}}{{\it {A neural Bayesian
  estimator for conditional probability densities}},
  }{}\href{http://arxiv.org/abs/physics/0402093}{{\tt
  arXiv:physics/0402093}}\relax
\mciteBstWouldAddEndPuncttrue
\mciteSetBstMidEndSepPunct{\mcitedefaultmidpunct}
{\mcitedefaultendpunct}{\mcitedefaultseppunct}\relax
\EndOfBibitem
\bibitem{Hulsbergen:2005pu}
W.~D. Hulsbergen, \ifthenelse{\boolean{articletitles}}{{\it {Decay chain
  fitting with a Kalman filter}},
  }{}\href{http://dx.doi.org/10.1016/j.nima.2005.06.078}{Nucl.\ Instrum.\
  Meth.\  {\bf A552} (2005) 566},
  \href{http://arxiv.org/abs/physics/0503191}{{\tt
  arXiv:physics/0503191}}\relax
\mciteBstWouldAddEndPuncttrue
\mciteSetBstMidEndSepPunct{\mcitedefaultmidpunct}
{\mcitedefaultendpunct}{\mcitedefaultseppunct}\relax
\EndOfBibitem
\bibitem{Aliev:2005np}
T.~M. Aliev and M.~Savci, \ifthenelse{\boolean{articletitles}}{{\it
  {Polarization effects in exclusive semileptonic $\Lb \rightarrow \Ls
  \ellp\ellm$ decay}},
  }{}\href{http://dx.doi.org/10.1088/1126-6708/2006/05/001}{JHEP {\bf 05}
  (2006) 001}, \href{http://arxiv.org/abs/hep-ph/0507324}{{\tt
  arXiv:hep-ph/0507324}}\relax
\mciteBstWouldAddEndPuncttrue
\mciteSetBstMidEndSepPunct{\mcitedefaultmidpunct}
{\mcitedefaultendpunct}{\mcitedefaultseppunct}\relax
\EndOfBibitem
\bibitem{Buras:1994dj}
A.~J. Buras and M.~Munz, \ifthenelse{\boolean{articletitles}}{{\it {Effective
  hamiltonian for $B \rightarrow X_s e^+ e^-$ beyond leading logarithms in the
  na\"ive dimensional regularization and~'t Hooft-Veltman schemes}},
  }{}\href{http://dx.doi.org/10.1103/PhysRevD.52.186}{Phys.\ Rev.\  {\bf D52}
  (1995) 186}, \href{http://arxiv.org/abs/hep-ph/9501281}{{\tt
  arXiv:hep-ph/9501281}}\relax
\mciteBstWouldAddEndPuncttrue
\mciteSetBstMidEndSepPunct{\mcitedefaultmidpunct}
{\mcitedefaultendpunct}{\mcitedefaultseppunct}\relax
\EndOfBibitem
\bibitem{Buras:1993xp}
A.~J. Buras, M.~Misiak, M.~Munz, and S.~Pokorski,
  \ifthenelse{\boolean{articletitles}}{{\it {Theoretical uncertainties and
  phenomenological aspects of $B \rightarrow X_s \gamma$ decay}},
  }{}\href{http://dx.doi.org/10.1016/0550-3213(94)90299-2}{Nucl.\ Phys.\  {\bf
  B424} (1994) 374}, \href{http://arxiv.org/abs/hep-ph/9311345}{{\tt
  arXiv:hep-ph/9311345}}\relax
\mciteBstWouldAddEndPuncttrue
\mciteSetBstMidEndSepPunct{\mcitedefaultmidpunct}
{\mcitedefaultendpunct}{\mcitedefaultseppunct}\relax
\EndOfBibitem
\bibitem{Hrivnac:1994jx}
J.~Hrivnac, R.~Lednicky, and M.~Smizanska,
  \ifthenelse{\boolean{articletitles}}{{\it {Feasibility of beauty baryon
  polarization measurement in the $\Ls \jpsi$ decay channel with the pp
  collider experiment}},
  }{}\href{http://dx.doi.org/10.1088/0954-3899/21/5/007}{J.\ Phys.\  {\bf G21}
  (1995) 629}, \href{http://arxiv.org/abs/hep-ph/9405231}{{\tt
  arXiv:hep-ph/9405231}}\relax
\mciteBstWouldAddEndPuncttrue
\mciteSetBstMidEndSepPunct{\mcitedefaultmidpunct}
{\mcitedefaultendpunct}{\mcitedefaultseppunct}\relax
\EndOfBibitem
\bibitem{LHCb-PAPER-2012-057}
LHCb collaboration, R.~Aaij {\em et~al.},
  \ifthenelse{\boolean{articletitles}}{{\it {Measurements of the $\Lb \to \Ls
  \jpsi$ decay amplitudes and the \Lb baryon production polarisation in $pp$
  collisions at $\sqrt{s} = 7\tev$}},
  }{}\href{http://arxiv.org/abs/1302.5578}{{\tt arXiv:1302.5578}}, {to appear
  in Phys.~Lett.~B}\relax
\mciteBstWouldAddEndPuncttrue
\mciteSetBstMidEndSepPunct{\mcitedefaultmidpunct}
{\mcitedefaultendpunct}{\mcitedefaultseppunct}\relax
\EndOfBibitem
\end{mcitethebibliography}
